\documentclass[conference]{IEEEtran}
\usepackage{listings}
\usepackage{caption}
\usepackage{xcolor}
\usepackage{bbding}
\usepackage[linesnumbered,ruled, noend]{algorithm2e}
\usepackage{enumitem}
\usepackage[utf8]{inputenc}
\usepackage{multirow}
\usepackage{array}
\usepackage{booktabs}
\usepackage{verbatim} 
\usepackage[tight,footnotesize]{subfigure}
\usepackage{algorithmic}
\usepackage[cmex10]{amsmath}
\usepackage{amssymb}
\usepackage{graphics}
\usepackage[pdftex]{graphicx}

\usepackage[hyphens]{url}

\ifCLASSINFOpdf
\graphicspath{{./pictures/}}
\DeclareGraphicsExtensions{.pdf,.jpeg,.png}
\fi

\definecolor{dkgreen}{rgb}{0,0.6,0}
\definecolor{gray}{rgb}{0.5,0.5,0.5}
\definecolor{mauve}{rgb}{0.58,0,0.82}
\definecolor{groovyblue}{HTML}{0000A0}
\definecolor{groovygreen}{HTML}{008000}
\definecolor{darkgray}{rgb}{.4,.4,.4}

\lstdefinelanguage{Groovy}[]{Java}{
	keywordstyle=\color{groovyblue}\bf,
	stringstyle=\color{mauve}\ttfamily,
	keywords=[3]{each, findAll, groupBy, collect, inject, eachWithIndex, subscribe, unsubscribe},
	morekeywords={def, as, in, use, Trigger, Condition, Action},
	moredelim=[is][\textcolor{darkgray}]{\%\%}{\%\%},
	moredelim=[il][\textcolor{darkgray}]{��}
}

\newcommand{\tool}{\textsc{HomeGuard}\xspace}
\newcommand{\tlong}{Cross-App Interference\xspace}
\newcommand{\tshort}{CAI\xspace}
\newcommand{\todo}[1]{{\textcolor{red}{#1}}}
\newcommand{\aaf}{\vspace*{-6pt}}

\begin{document}
\title{\tlong Threats in Smart Homes:\\Categorization, Detection and Handling}

\author{\IEEEauthorblockN{Haotian Chi\IEEEauthorrefmark{1}, Qiang Zeng\IEEEauthorrefmark{2}, Xiaojiang Du\IEEEauthorrefmark{1}, Jiaping Yu\IEEEauthorrefmark{1}}
	\IEEEauthorblockA{\IEEEauthorrefmark{1}Department of Computer and Information Sciences, Temple University, Philadelphia, PA 19122, USA \\
	\IEEEauthorrefmark{2}Department of Computer Science and Engineering, University of South Carolina, Columbia, SC 29208, USA}
	Email: \{htchi, dux, jiaping.yu\}@temple.edu, Zeng1@cse.sc.edu}


%

\IEEEoverridecommandlockouts
\makeatletter\def\@IEEEpubidpullup{5\baselineskip}\makeatother
\IEEEpubid{\parbox{\columnwidth}{An earlier version of this paper was submitted to ACM CCS'18 on May 9th, 2018. This version contains some minor modifications based on that submission.
	}
	\hspace{\columnsep}\makebox[\columnwidth]{}}
\maketitle

\maketitle

\begin{abstract}
A number of Internet of Things (IoTs) platforms have emerged to enable various 
IoT apps developed by third-party developers to automate smart homes. 
Prior research mostly concerns the overprivilege problem in the permission model.
Our work, however, reveals that even IoT apps that follow the principle of least
privilege, when they \emph{interplay}, can cause unique types of threats,
named \emph{\tlong} (\tshort) threats. We describe and 
categorize the new threats, showing that unexpected automation, security and 
privacy issues may be caused by such threats, which cannot be handled
by existing IoT security mechanisms. To address this problem, 
we present \tool, a system for appified IoT platforms 
to detect and cope with \tshort threats. A symbolic executor module is built to 
precisely extract the automation semantics from IoT apps. The semantics of different IoT 
apps are then considered collectively to evaluate their interplay  and discover 
\tshort threats systematically. A user interface is presented
to users during IoT app installation, interpreting the discovered threats to help
them make decisions. We evaluate \tool via a proof-of-concept implementation on Samsung's SmartThings and discover many threat instances among apps in the SmartThings public repository.
The evaluation shows that it is precise, effective and efficient.
\end{abstract}

\section{Introduction}
The rapid proliferation of Internet-of-Things (IoTs) has advanced the 
development of smart homes to a new era where emerging centralized 
and appified (a.k.a., app-powered) smart home platforms connect heterogenous IoT devices and offer 
programming frameworks for third-party developers to contribute 
various home automation ideas. According to the report of German 
IoT market research firm IoT Analytics in July 2017, the number 
of IoT platforms on the market has grown from 360 to 450 over 
the past 12 months and smart home platforms account for 21 percent \cite{iot2017platform}. 
Representative examples of such platforms are Samsung SmartThings \cite{smartthings2018}, 
Apple HomeKit \cite{homekit2018}, and Google Android Things \cite{andoirdthings2018}. 
By installing IoT apps on a platform, users can monitor, remotely control, 
and automate their IoT devices to make smarter homes. However, appified 
IoT platforms also introduce new app-level attack surfaces, which can be 
exploited by attackers or misused by homeowners, introducing new challenges 
in security and privacy. For example, burglars can stealthily open a 
smart door lock via an IoT app to break into homes, which is impossible 
in non-appified IoT systems.

Fernandes et al.~\cite{fernandes2016security} discover design flaws such 
as the \emph{overprivilege} problem in Samsung's SmartThings, 
one of the most mature smart home 
platforms; they demonstrate that malicious apps can be constructed to expose 
smart homes to severe attacks that exploit the overprivilege problem. 
Thus, some systems are proposed to handle the  problem. 
ContexIoT \cite{jia2017contexiot} proposes a context-based permission system 
to involve users into making decisions on whether a security-critical command 
in an IoT app during runtime should proceed under the current context. SmartAuth \cite{tian2017smartauth} 
compares the analysis result of the app code with the code annotations and app description
to identify overprivilege in an app, and allows users to specify access control policies.  
Tyche \cite{rahmati2018tyche} designs a risk-based permission model over the 
original model which groups all supported permission-defined commands into three risk levels, 
and allows users to specify a maximum risk level for any installed app. 
Thus, permissions greater than the specified level will be denied during the app execution.
Instead of handling security threats proactively, ProvThings \cite{wang2018fear} presents 
a runtime logging system  for forensics and diagnosis purposes.

This paper reveals that new types of threats, which we call \emph{\tlong} (\tshort) threats, can be
caused by apps even if the permission system strictly follows the principle of least privilege.
Hence, \tshort threats do not depend on the overprivilege problem;
when IoT apps installed at the same smart environment \emph{interplay}, various \tshort threats may arise.

First, when
a smart home is installed with apps provided by different third-party developers,
there is a chance that some apps may contain \emph{conflicting} logics in terms of 
how to control an IoT device. For example, given a certain combination of
sensor values, two apps issue opposite commands for opening/closing
a smart door, which not only confuses homeowners but also causes 
seriously exploitable security issues. 

Second, the action taken by
a smart device, due to the execution of one app, may trigger a \emph{chained} execution
of another app, but not all such chained executions are desired by homeowners 
and some may even cause security or privacy issues. For example, assuming
an app is to turn off the light when no motion is detected for a period of time, while
another app is to open the curtain if the room is \emph{too} dark during the daytime,
then opening the curtain becomes a chained action due to the light-off action.
Such chained actions may occur only under specific circumstances; thus,
they impose security and privacy threats that are hidden from users. 

In addition to threats imposed by benign apps, attackers may construct \tshort
threats by including seemingly benign logic into a malicious app to 
successfully pass the code review. But the app is to interact with other 
apps to launch attacks, e.g., opening the smart door. That is, exploitation of
such threats can become a \emph{new attack vector} against smart environments.

Therefore, it is important and urgent to present, detect, and handle the new types of threats.
The goal of this work is thus to (1) describe \tshort threats and categorize them,
(2) propose and implement a technique to automatically discover such 
threats, and (3) present the revealed threats to homeowners in
a user-friendly way and allow them to make informed decisions in handling these threats. 

Existing techniques cannot discover \tshort threats, as they analyze
apps individually and most of them focus on IoT app-level threats exploiting overly granted permissions, 
while \tshort threats are essentially due to
the intricate interplay between multiple apps. Hence, discovering the new type of threats calls for \emph{cross-app-boundary} semantics-relation analysis
(i.e., how the logic defined in one IoT app may interfere with that of another app),
which is the main challenge of our work. 

Our idea is to extract the automation semantics (also referred to as \emph{rules}) from each app and then
discover the threats by evaluating the interaction relations between the extracted
semantics collectively and systematically. To precisely capture the semantics of an
app, we propose to perform symbolic execution analysis on IoT apps. 
The semantics of each app is then represented as quantifier-free
first-order formulas. Lastly, an automatic method based on constraint solver 
is applied for discovering \tshort threats, which are then interpreted into a human-readable
form and presented to homeowners.   
 
Instead of requiring the repetitive
intervention of users to handle such threats, we propose a deployment method that is
much more user-friendly and easy to deploy. Whenever a new app
is to be installed, our system interposes and analyzes whether
there exist \tshort threats between the new app and the already-installed ones. 
As a result, the homeowner only needs to
spend one-time decision making for each app to be installed. 

We develop a proof-of-concept prototype system \tool on Samsung SmartThings, 
which at the time of research supports the largest number of IoT devices 
and IoT apps (i.e., SmartApps). Our evaluation shows that \tool can precisely discover
\tshort threats from real-world SmartApps, and generate the
analysis results instantly.  

Our main contributions can be summarized as follows:
\begin{itemize}[leftmargin=*]
	\item We describe new types of threats in smart environments, \emph{\tlong} 
threats, which do not depend on overprivileged apps. Attackers may exploit such threats
by inserting seemingly benign logics so that malicious apps can pass the conventional code 
review but may interact with other apps in a harmful way. Besides, even benign
apps, due to their interplay, may cause undesired or even dangerous consequences.

	\item We present a semantics-based user-friendly system \tool that 
not only discovers \tshort threats but also assists users to handle them.
 Our approach introduces zero-modification to the smart home platform architecture  
and only needs user intervention during app installation. 

	\item We design and build a symbolic execution engine that can precisely 
extract automation semantics from SmartApps. To our knowledge, this is the first
symbolic executor for SmartApp code. 

	\item  We evaluate the effectiveness and efficiency of \tool. 
The evaluation shows that it can detect all of the categories of \tshort threats
precisely and efficiently. Our experiments discover and verify 
many threat instances from real-world apps in the SmartThings public repository. 
\end{itemize}

\section{Background}
We first describe the home automation model, and then
introduce the popular cloud-backed smart home platforms. 


\begin{figure}[tb]
	\centering
	\includegraphics[width=0.48\textwidth]{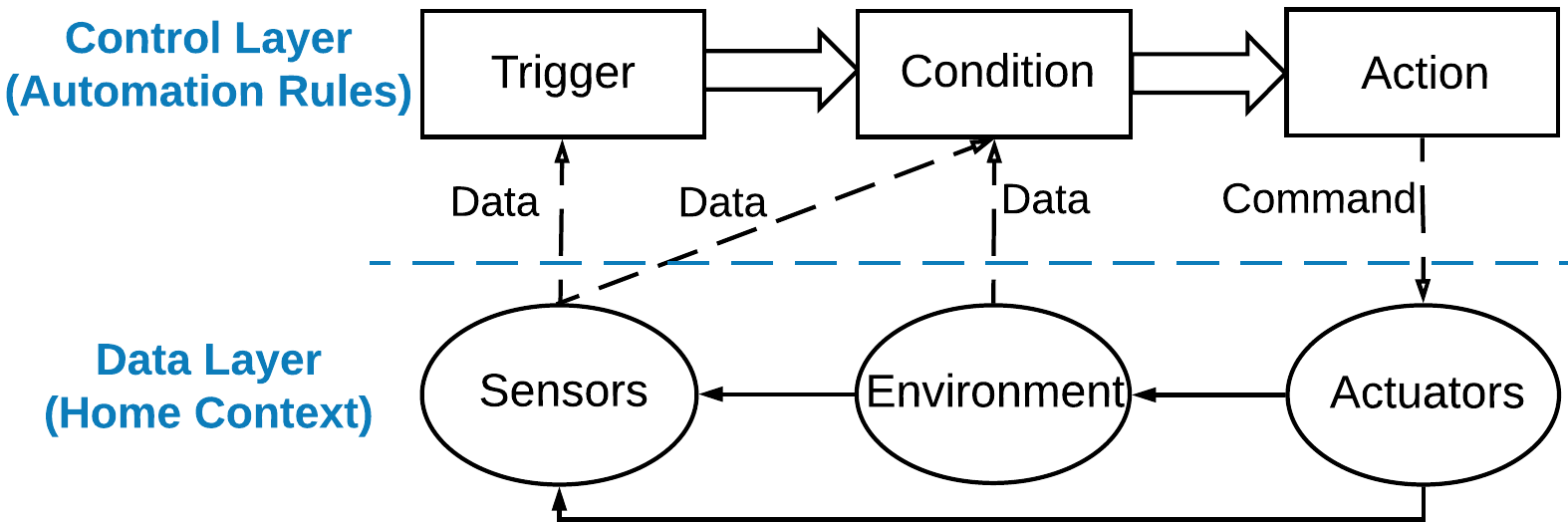}\\
	\caption{The home automation model.}\label{fig_tca}
\end{figure}

\subsection{Home Automation Model} \label{sec:home-model}
A home automation model can be abstracted into the \emph{data layer} and the \emph{control layer}. 

\vspace{2pt}
\noindent\textbf{Data Layer.} The data layer of a home is its \emph{home context}, consisting 
of sensors, actuators, and the home environment. (1) A \emph{sensor} in our model 
may be a traditional device or subsystem that measures a specific feature in the home
environment (e.g., the temperature), a device that can report a certain attribute of 
itself (e.g., a switch can report its \texttt{on}/\texttt{off} status), 
or a platform-defined feature (e.g., a location mode in SmartThings). 
(2) An actuator can be either a controllable device or platform-defined feature. 
An IoT \emph{device} may be a sensor, an actuator or a combination.
(3) The home environment has a set of measurable natural features, including time, 
temperature, illuminance, humidity, power consumption, etc. 
The interaction relation of sensors, actuators, and the environment is shown 
in Fig.~\ref{fig_tca}. Sensors observe the home 
environment features and output the corresponding measurements while
actuators can influence the reading of sensors either directly 
(e.g., by altering its own device attribute) or via the environment 
(e.g., by heating the home to change the measurement reading of a temperature sensor).

\vspace{2pt}
\noindent\textbf{Control Layer.}
In appified home automation systems, the control layer consists of the automation \emph{rules} defined by home automation apps. 
An app usually defines one or more rules.
The rule follows a \emph{trigger-condition-action} (TCA) model, as depicted in Fig.~\ref{fig_tca}: 
(1) The \emph{trigger}  
is a subscribed event (e.g., the state of television changes to \texttt{on}) that fires the execution of the remaining flow of the rule. 
The platform listens to all data reported by sensors, and then broadcasts the 
related events to the subscribers.  
(2) The \emph{condition} is a set of 
constraints defined in terms of the sensor data, the environment data (e.g., the time) or a user 
input (e.g., the room temperature is above 30$^{\circ}$C), which must be satisfied 
in order to take the action. 
(3) The \emph{action} (e.g., to open the window) 
is one or more commands issued to the actuators. 

The data layer and the control layer interact in both directions.
On the one hand, rules obtain
data from the home context. For example, the trigger of a rule is usually  a 
specific sensor data update (e.g., the state of TV changes 
to \texttt{on}); the condition takes sensor data and environment measurements as the inputs 
for constraint check.  On the other hand, rules can influence the home context, 
i.e., the actions issue commands to the actuators, which may, in turn, change the environment 
features and sensor readings.

\begin{figure}[tb]
	\centering
	\includegraphics[width=0.48\textwidth]{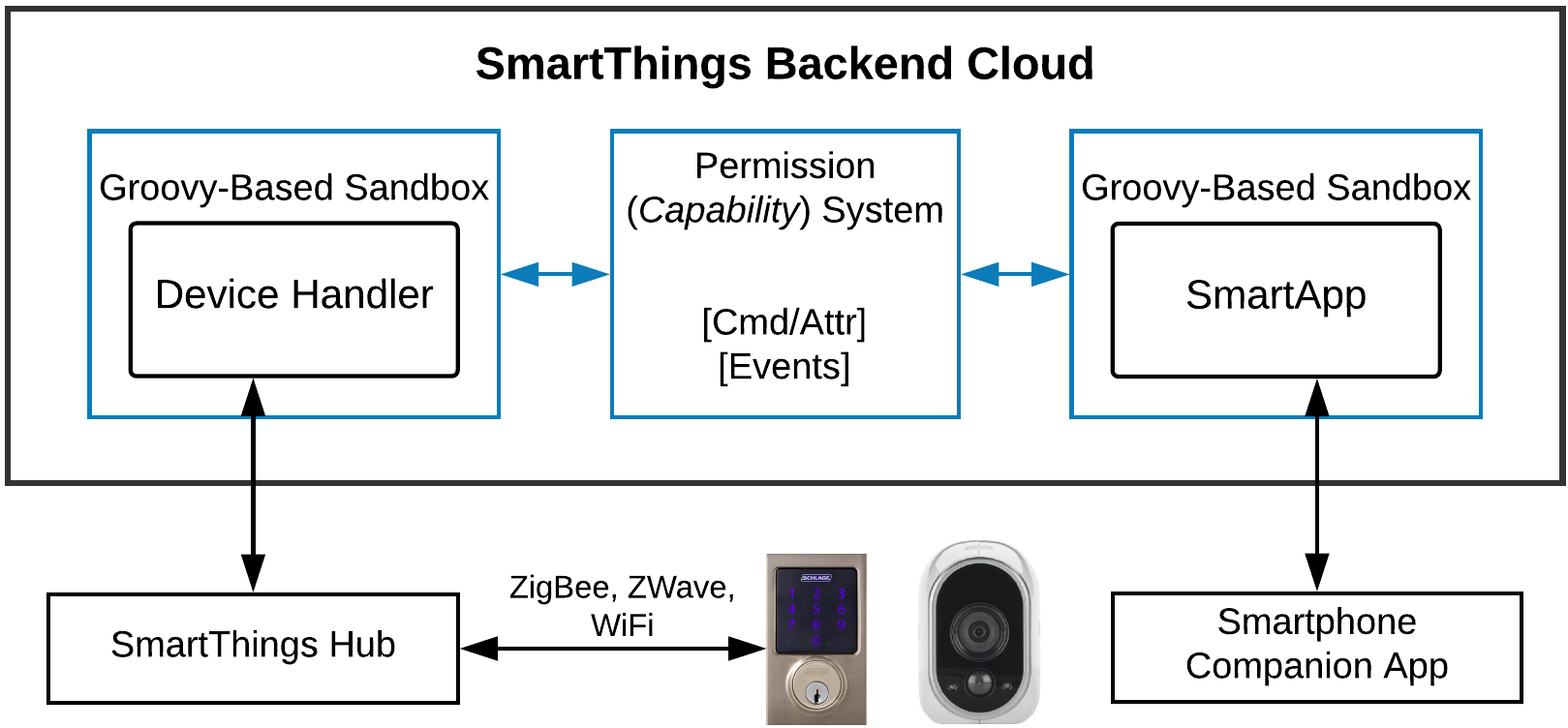}\\
	\caption{SmartThings architecture overview.}\label{fig_smartthings}
\end{figure}

\subsection{Appified Smart Home Platforms}
Recently, multiple cloud-backed smart home platforms have emerged.
They allow third-party developers to publish their IoT apps to help users manage, 
monitor and control their home devices. The integration of app developers is significant 
to share novel ideas and build smarter homes.
A typical cloud-backed smart home ecosystem comprises three necessary components: 
a \emph{hub} connecting IoT devices, a \emph{back-end cloud}, and a \emph{smartphone app}. 

We use the Samsung SmartThings platform (as shown in Fig.~\ref{fig_smartthings}), 
one of the most popular smart home platforms, as an example to describe these components.
(1) The hub connects all IoT devices through 
short-range wireless techniques (ZigBee, Z-Wave, Bluetooth) or WiFi and bridges the communication between IoT devices and the Internet. Each of the low-cost and 
resource-constrained IoT devices usually only supports one of these wireless access techniques, 
while a hub is equipped with most, if not all, of the popular wireless capabilities. The hub plays 
a core role to ensure the interconnectivity and interoperability of IoT devices in 
the home environment. (2) The backend cloud (also referred to as \emph{platform} in this paper), 
is another key part in the system. The SmartThings cloud abstracts real devices to \emph{device handler} instances, which are software wrappers of the physical 
devices. A device handler handles the real underlying communication between the cloud and a physical 
device and exposes a subset of pre-defined uniform interfaces to the other 
components within the platform. \emph{SmartApps} are Groovy programs provided by third-party 
developers and are used to provide automation rules for the home devices. Specifically, 
SmartApps can subscribe to the \emph{events} fired by a set of instances of device handlers
and issue \emph{commands} to device handlers; thus, SmartApps interact with the real devices indirectly. The SmartThings provide the sandboxed runtime environment for running SmartApp and device handlers.
(3) To provide a convenient user interface (UI) for users to manage their hubs, IoT devices 
and SmartApps, SmartThings also provide a smartphone \emph{companion} app with a SmartApp store. 
In the companion app, users can install and configure a SmartApp, e.g.,
granting IoT devices that support the \emph{capabilities}\footnote{See Appendix~\ref{appendix_smartthings_api_sink} for more details about capabilities.} requested by SmartApps, and filling out 
user-defined values. We call user-provided information in this phase the 
\emph{configuration information}.

\section{Categorization of \tshort Threats}
\label{section_attacks}

\begin{table*}[t]
	\centering
	\footnotesize
	\caption{Categorization of \tshort threats. 
		${R}_{i}=(T_{i}, C_{i}, A_{i}), i=1, 2$ denotes two rules, where $T_{i}$, $C_{i}$, $A_{i}$ are the
		trigger, condition and action, respectively. $A_{i}\mapsto T_{j}$ denotes $A_{i}$ satisfies
		$T_{j}$. $A_{i}\Rightarrow C_{j}$ and $A_{i}\nRightarrow C_{j}$ denotes that $A_{i}$
		satisfies and dissatisfies a subset of constraints in $C_{j}$, respectively. $G(A_{i})$ denotes the goal of $A_i$.}
	\label{table_attacks}
	\renewcommand\arraystretch{1.41}
	\newcommand{\tabincell}[2]{\begin{tabular}{@{}#1@{}}#2\end{tabular}}
	\newcolumntype{P}[1]{>{\centering\arraybackslash}p{#1}}
	\newcolumntype{M}[1]{>{\centering\arraybackslash}m{#1}}
	\begin{tabular}{|M{2.5cm}|p{9.1cm}|p{5.4cm}| }		
		\hline
		 & Category and description  & Pattern \\\hline	 
		\multirow{2}*{\tabincell{c}{Action-Interference \\Threats}} &	\textbf{Actuator Race (AR)}: Two rules perform contradictory actions on the same actuator(s).  & \scriptsize\tabincell{c}{$T_{1}=T_{2}$, $C_{1}\cap C_{2}\neq \emptyset$, $A_{1}=\neg A_{2}$}   \\
		~ &	\textbf{Goal Conflict (GC)}: Two rules' actions have contradictory goals.  & \scriptsize\tabincell{l}{($T_{1}\cup C_{1})\cap (T_{2}\cup C_{2})\neq \emptyset$, $G(A_{1})=\neg G(A_{2})$}  \\
		\hline
		\multirow{3}*{\tabincell{c}{Trigger-Interference \\Threats}} &	\textbf{Covert Triggering (CT)}: A rule triggers the execution of other rules. & \scriptsize\tabincell{l}{$A_{1}\mapsto T_{2}$, $C_{1}\cap C_{2}\neq \emptyset$}  \\
		~& \textbf{Self Disabling (SD)}: A rule triggers other rules which in turn disables it.  & \scriptsize\tabincell{l}{$A_{1}\mapsto T_{2}$, $C_{1}\cap C_{2}\neq \emptyset$, $A_{2}=\neg A_{1}$} \\
		~ & \textbf{Loop Triggering (LT)}: Two rules trigger each other but perform contradictory actions on the same actuator(s). & \scriptsize\tabincell{l}{$A_{1}\mapsto T_{2}$,  $A_{2}\mapsto T_{1}$, $C_{1}\cap C_{2}\neq \emptyset$}, $A_{1}=\neg A_{2}$ \\
		\hline
		
		\multirow{2}*{\tabincell{c}{Condition-Interference \\Threats}} &\textbf{Enabling-Condition (EC)}: A rule enables the execution of other rules.   & \scriptsize\tabincell{l}{$A_{1}\Rightarrow C_{2}$} \\
		~ &\textbf{Disabling-Condition (DC)}: A rule disables the execution of other rules.   & \scriptsize\tabincell{l}{$A_{1}\nRightarrow C_{2}$} \\
		\hline
		
	\end{tabular}
	\vspace{3mm}
\end{table*}

A rule can be modeled as a tuple of ``\emph{trigger-condition-action}''. Suppose $R_{1}=(T_{1}, C_{1}, A_{1})$ and $R_{2}=(T_{2}, C_{2}, A_{2})$ denote two rules that are installed in the same environment, where $T_{i}$, $C_{i}$, $A_{i}$ are the trigger, condition, and action of $R_{i}$, respectively.
A \tlong threat arises when \emph{the action of $R_1$ interferes 
with the trigger, condition, or action of $R_2$.}
$R_1$ and $R_2$ may or may not belong to the same app,
and our system \tool can handle both cases, so we do not distinguish between the two cases in this paper. 

Based on how $R_2$ is interfered with by $R_1$,
we have identified the following three \emph{basic} classes of \tshort threats:
\emph{Trigger-Interference Threats}, \emph{Condition-Interference Threats},
and \emph{Action-Interference Threats}, which arise when 
the trigger, condition, and action of $R_2$ is interfered with by
the action of $R_1$, respectively. We summarize the different categories of \tshort threats in Table \ref{table_attacks}. The caption of Table \ref{table_attacks} gives the notations in this section.

\begin{figure}[t]
	\centering
	\includegraphics[width=0.43\textwidth]{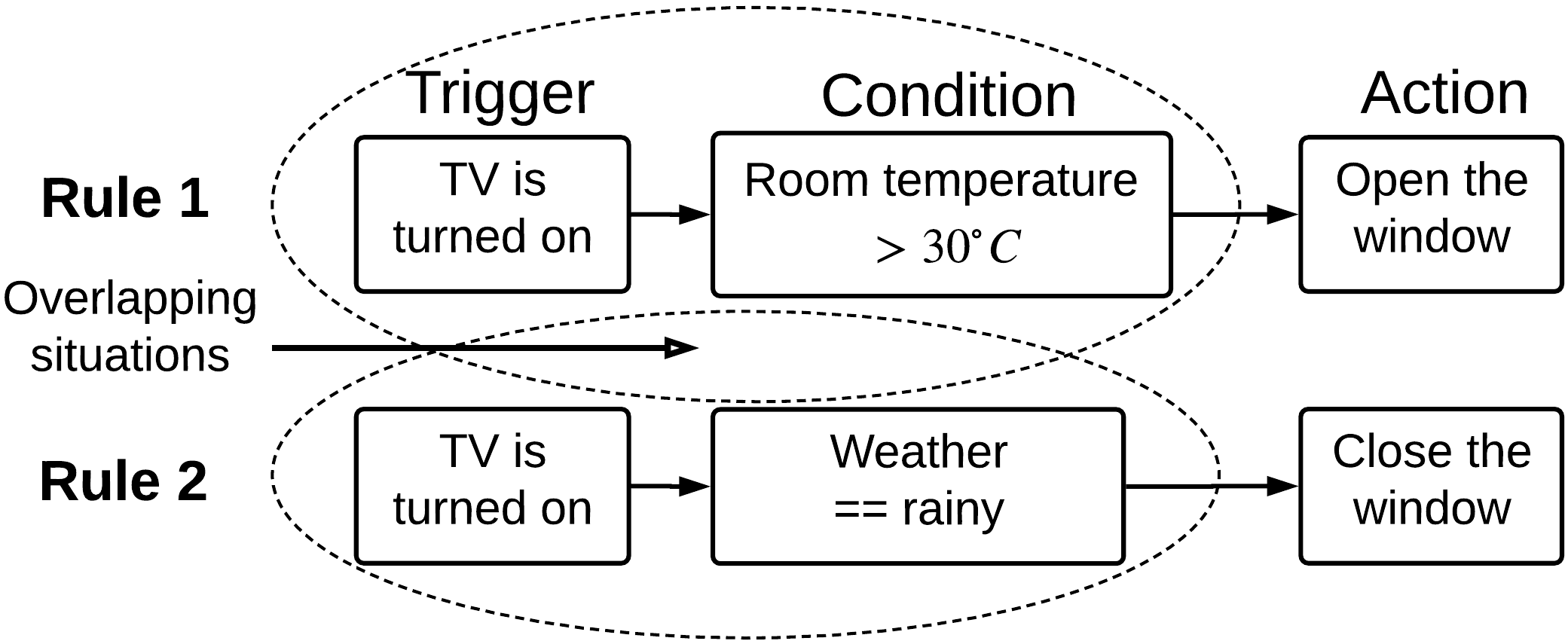}\\
	\caption{An example of Action-Interference Threats. }\label{fig_action_conflicts}
\end{figure}

\subsection{Action-Interference Threats}
Two rules $R_1$ and $R_2$ may be programmed to operate on the same
actuator under different circumstances (decided by their own triggers and conditions).
It is likely that both rules take effect simultaneously if they are both triggered ($T_{1}=T_{2}$) and their conditions are both satisfied ($C_{1}\cap C_{2}\neq\emptyset$). 
If $R_1$ and $R_2$ issue contradictory commands to the same actuator ($A_1=\neg A_2$), 
an \emph{Actuator Race} threat may arise, such that the final status of the actuator becomes unpredicted. 

Fig.~\ref{fig_action_conflicts} shows such an example. 
If it is raining and the temperature is above 30$^{\circ}$C, the 
conditions of both rules are satisfied, and both rules are triggered when the TV is turned on. 
However, the two rules issue opposite commands on the window, leading to a race condition. 
To verify the threat, we have constructed two SmartApps that define \textbf{Rule 1} 
and \textbf{Rule 2} shown in Fig.~\ref{fig_action_conflicts}, respectively, 
and run them to control the same window opener's switch. We observed a variety of results: 
the switch is turned on only, turned off only, turned on then off, and turned off then on,
showing that the final state is unpredictable. 

Note that such races arise only under specific conditions. 
As a result, the threats may hide in a smart environment for a prolonged time.
Such races may confuse the user and cause annoyance or device damages; they may
even be exploited by attackers (e.g., the window is left open without being expected by the user).


A variant of Action-Interference Threats arise when two rules are executed on different actuators in certain situations ($T_{1}\cup C_{1})\cap (T_{2}\cup C_{2})\neq\emptyset$), 
but their effects contradict ($G(A_{1})=\neg G(A_{2})$), which we call \emph{Goal Conflicts}. Different from the instant race on an actuator, the two rules do not need to be triggered at the same time. The inter-actuator conflict is 
more subtle and implicit than intra-actuator races and hence more difficult to be realized by users. 
For instance, one rule is to turn on a heater, while the other is to open the window if the room is too dark;
the two actions conflict in terms of heating up the room. 


\begin{figure}[t]
	\centering
	\includegraphics[width=0.43\textwidth]{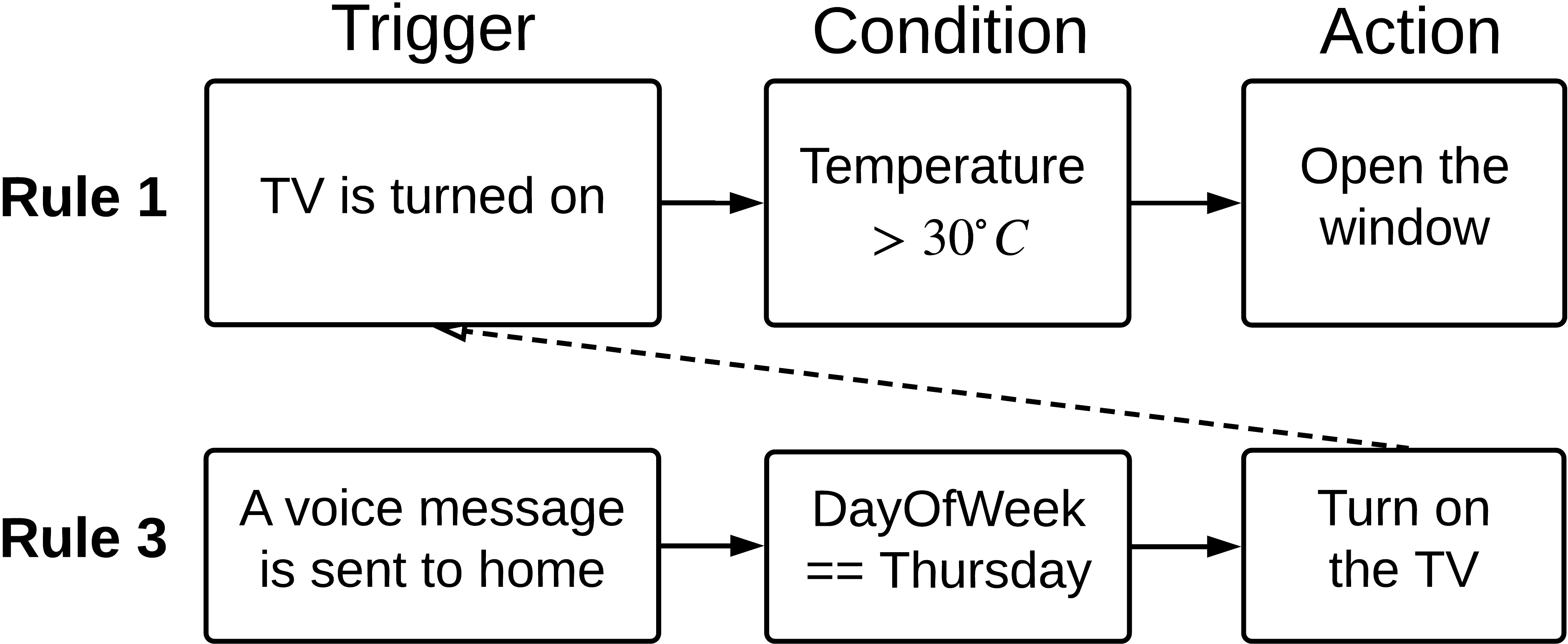}\\
	\caption{An example of Trigger-Interference Threats.}\label{fig_covert_rules}
	\aaf\aaf
\end{figure}

\subsection{Trigger-Interference Threats}
\label{section_trigger_interference}
Since all rules interact within a common home context, a rule $R_{1}$'s action may change 
the home context to a status that triggers another rule $R_{2}$ ($A_{1}\mapsto T_{2}$); if the triggering happens under certain circumstances where $R_{2}$'s condition is satisfied ($C_{1}\cap C_{2}\neq \emptyset$), $R_{2}$ is also executed after $R_{1}$.
Thus, a group of explicitly defined rules may lead to new implicit rules, which
we call \emph{covert rules}. As shown in Fig.~\ref{fig_covert_rules}, \textbf{Rule 3} 
turns on the TV, which then triggers the execution of \textbf{Rule 1}. 
A covert rule ``\emph{\textbf{when} sending a voice message \textbf{if} it is on Thursday and the temperature 
is over 30$^\circ$C \textbf{then} open the window}'' is formed. 

In this example, a user may only 
intend to use \textbf{Rule 3} to turn on the TV remotely for the purpose of watching a live 
show immediately when arriving home, but it also triggers \textbf{Rule 1} to 
open the window, creating a chance for a burglar to break in before the user arrives home.
In other words, such \emph{Covert Triggering} may not be desired by the users. Thus, it is
important to find such covert rules and alert the users when apps are installed.

There are two special cases of Trigger-Interference Threats: (1) \emph{Self Disabling} that happens when $R_{1}$ covertly triggers $R_{2}$ but $R_{2}$'s action contradicts that of $R_{1}$ ($A_{2}=\neg A_{1}$). For instance, $R_{1}$ is defined as ``\emph{\textbf{when} the motion sensor detects a motion at the front door \textbf{if} the temperature is above 30$^{\circ}$C \textbf{then} turn on the air conditioner}'', and $R_{2}$ is ``\emph{\textbf{when} the energy meter's reading exceeds a threshold \textbf{then} turn off the air conditioner}''. If turning on the air conditioner makes the reading of the power meter exceeds a threshold, the air conditioner will be turned off immediately after it is turned on. (2) \emph{Loop Triggering} that occurs when two rules trigger each other ($A_{1}\mapsto T_{2}$,  $A_{2}\mapsto T_{1}$, $C_{1}\cap C_{2}\neq \emptyset$) but perform contradictory actions on the same actuator(s) ($A_{1}=\neg A_{2}$). For example, $R_1$ is defined as  ``\emph{\textbf{when} the illuminance is below 30 LUX \textbf{then} 
turn on the lights}'', and $R_2$ is ``\emph{\textbf{when} the illuminance is 
above 50 LUX \textbf{then} turn off the lights}''; as a result, when the two rules
control the same set of lights, the lights may be continuously turned on and off,
which can not only cause device damages, but also lead to seizures in photosensitive 
epilepsy sufferers \cite{ronen2016extended}.

\begin{figure}[t]
	\centering
	\includegraphics[width=0.43\textwidth]{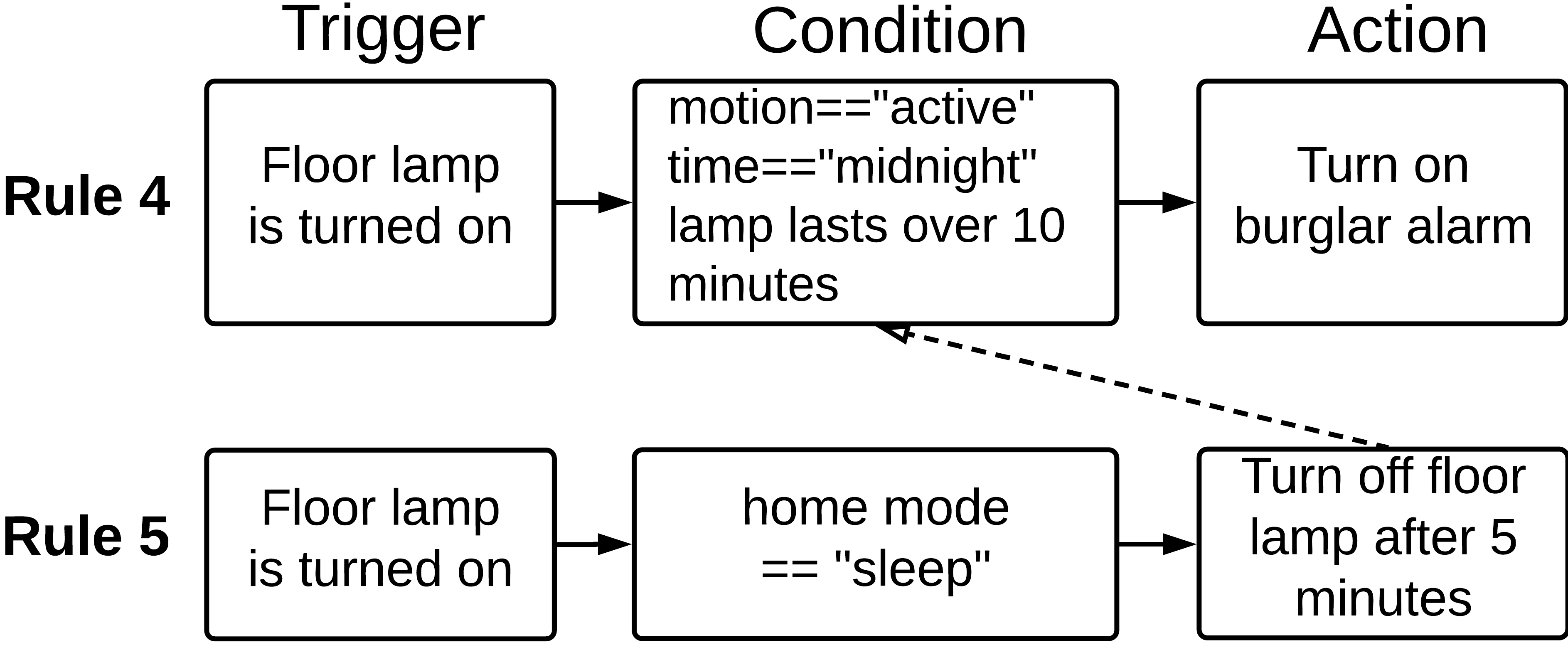}\\
	\caption{An example of Condition-Interference Threats.}\label{fig_condition_interference_threats}
	\aaf\aaf
\end{figure}
\subsection{Condition-Interference Threats}

The action due to a rule $R_1$ may change the satisfaction of another rule $R_2$'s condition,
and thus affect the execution of $R_2$, which are called Condition-Interference Threats. 
However, unlike Trigger-Interference Threats, the action of $R_1$ cannot directly trigger
the execution of $R_2$, as $R_2$ has its own trigger.

There are two types of Condition-Interference Threats: \emph{Enabling-Condition Interference}
and \emph{Disabling-Condition Interference}, depending on whether the action of $R_1$ changes the condition of $R_2$ from
\texttt{false} to \texttt{true} (Enabling, $A_{1}\Rightarrow C_{2}$) or from \texttt{true} to \texttt{false} (Disabling, $A_{1}\nRightarrow C_{2}$).
Fig.~\ref{fig_condition_interference_threats} shows an example of Disabling-Condition Interference Threats,
where \textbf{Rule 5} turns off the floor lamp automatically to save energy when the 
home is in the ``\texttt{sleep}" mode, while \textbf{Rule 4} is used to detect break-ins. 
The action ``turning off the floor lamp" in \textbf{Rule 5} disables the floor 
lamp checking in \textbf{Rule 4} and may lead to false negatives.

\section{Problem Scope and System Overview}
\subsection{Threat Model}
\label{section_threat_model}
\tlong threats do not depend on the overprivilege 
problem of the smart things platform; moreover, they do not rely
on malicious code to be inserted into a \emph{single} app, which may not be
able to pass the manual code review step (e.g., code review of each submitted 
app is enforced by Samsung SmartThings). Thus, \tshort threats are more stealthy and 
cannot be handled by existing approaches that analyze or review  apps individually.

Our defense system considers and addresses \tshort threats that can be caused by
three types of IoT apps:
\begin{itemize}[leftmargin=*]
	\item \textbf{Flawless Benign Apps}: Although they contain no malicious 
or flawed code, they may cause \tshort threats because of interaction with other apps.
	\item \textbf{Flawed Benign Apps}: They contain flawed logic, e.g.,
inconsistent rules, that leads to \tshort threats. 
	\item \textbf{Malicious Apps}: They contain malicious code that purposely
exploits other apps deployed in the same home. The attacker can submit malicious 
apps onto the app store, or trick users to install them through the web. 
\end{itemize}

Even non-malicious apps can cause \tshort threats, which can surprise and confuse
users and lead to security and privacy issues without involving any attackers. 
Although users may sometimes perceive \tshort threats and avoid installing risky apps, 
various reasons lead to the failure of identifying such threats by relying on the carefulness of ordinary users. First, users may not understand
all functionalities of an app by only reading its description. Second, a home may have multiple users; even a single user may install an app first, and after a long time, install another
controlling the same device for distinct purposes. Third, users may write their own apps or obtain apps from third-party sources to realize novel automation ideas (e.g., controlling devices for a special purpose, achieving the same functionality with the limited devices they own, etc.). However, these apps may be flawed and vulnerable to \tshort threats when working with other apps. Therefore, an automatic detection technique is necessary.

\vspace{2pt}
\noindent\textbf{New Attack Vectors.}
Moreover, we identify three new attack vectors that exploit \tshort threats. 
(1) If an attacker can infer or obtain the information about the apps installed at
a target home, he can find the \tshort threats at this location
(e.g., the window is opened at a specific time) and exploit them accordingly.
(2) The attacker can publish seemingly benign but malicious apps onto the app store, or
trick users into installing them by advertising some useful functionalities.
Such apps can be built to cause \tshort threats by taking advantage
of other apps installed at the target home.
(3) If multiple collusive malicious apps are installed at
the same home, they can cooperate to construct \tshort threats and launch
powerful finely-controlled attacks.

\subsection{Goal and Problem Scope} 
\label{section_goal_scope}
The goal of our system is to detect \tshort threats in a smart home,
no matter they are caused by benign or malicious apps. 
This paper broadly uses \emph{threats} to
refer to all discovered interferences, acknowledging that \emph{some of them may
be unexpected and dangerous while others may be desired by users}.
Instead of distinguishing the two cases, which is probably a non-computable problem,
we present the detection results to homeowners in a human-readable
way, which alerts the owners and allows them to make decisions
on whether or not to keep the new app and or re-configure it.

In this paper, we focus on application security with the assumption of trusted 
and uncompromised platforms, devices and communication protocols. For example, 
attacks that exploit the hardware vulnerabilities or communication protocol 
flaws to intercept or fake the data used by an app and thus lead to an 
incorrect app behavior are out of our scope and should be taken care of by the 
device manufacturers, protocol designers and IoT platform to guarantee the 
authenticity and integrity of the data.

\subsection{System Overview}

\begin{figure}[tb]
	\centering
	\includegraphics[width=0.46\textwidth]{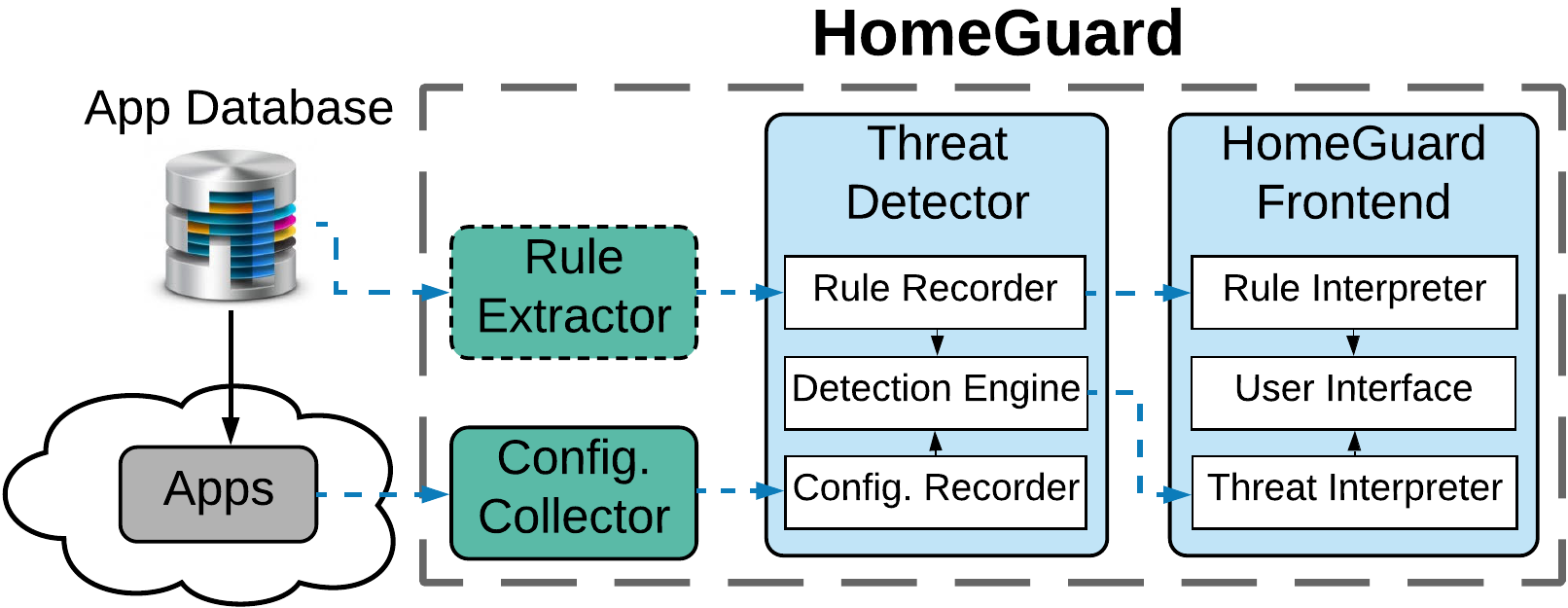}\\
	\caption{The architecture of \tool.}\label{fig_architecture}
	\vspace{-1em}
\end{figure}

In order to detect the \tshort threats in a given home, all the apps and their 
interaction need to be considered collectively and systematically. 
To speed up the detection, the design of \tool has the \emph{offline} part 
and the \emph{online} part: the offline part extracts and represents the automation semantics
of each app in the form of rules (Section~\ref{sec:home-model}); when a new app is being installed, the online
part is invoked to detect \tshort threats by considering the interplay of all
the rules of the new app and installed apps. The detection result is then presented
to users for decision making. The advantage is that the online part
can be computed efficiently by making use of the rules extracted offline. 
We build \tool using a modular design.
As shown in Fig.~\ref{fig_architecture}, it comprises the following four modules:

\begin{itemize}[leftmargin=*]
	\item \textbf{Rule extractor} extracts the rules of each smart home app
offline. It exposes APIs for querying the rules 
of an app and also provides online rule extraction services for users who use custom apps. 
Other modules of \tool are per-user processes and are called each time 
a new app $\mathcal{A}$ is being installed.

	\item \textbf{Configuration collector} collects the \emph{configuration information} 
related to the installation, which includes the app name, the devices bound 
to the app $\mathcal{A}$, and the in-app static values (thresholds, boolean values, contact 
information, etc.).

	\item \textbf{Threat detector} is invoked when the \emph{configuration recorder} 
receives the configuration information from \emph{configuration collector}. After that, 
the \emph{rule recorder} requests the rule information of  $\mathcal{A}$ from the \emph{rule extractor}. 
The \emph{rule recorder} and the \emph{configuration recorder} keep track of the historical 
rule and configuration information of apps, respectively. Based on the newly received rule and 
configuration information and the historical records, the \emph{detection engine} detects 
\tshort threats by considering the interplay of $\mathcal{A}$ and already installed apps. 

	\item \textbf{\tool frontend} bridges the system and smart home users. 
The \emph{rule interpreter} translates rules of $\mathcal{A}$ into a human-readable 
form and displays them via a \emph{user interface}, such that users can check if $\mathcal{A}$ 
itself will behave as it claims. The \emph{threat interpreter} displays the 
detected \tshort threats to users in a readable manner, who then decide whether the installation should proceed or whether the configuration should be adjusted.

\end{itemize}

Note that the \emph{rule extractor} and the \emph{configuration collector} are 
platform-specific since different IoT platforms use different programming languages 
and provide different APIs, while the \emph{threat detector} and the \emph{\tool frontend}
are platform-independent. The details of the \emph{rule extractor} and the \emph{threat detector} 
will be presented in Section~\ref{section_rule_extraction} and Section~\ref{section_threat_detector},
respectively. We present the other modules in Section~\ref{section_implementation}.

\section{Rule Representation and Extraction}
\label{section_rule_extraction}

\begin{lstlisting}[caption={Code snippet of \texttt{ComfortTV}. Some irrelevant lines (e.g., metadata definition, UI-related sections and pages) are omitted.},label=listing_rule1,float=tp,floatplacement=tbp, basicstyle=\scriptsize]
input "tv1", "capability.switch", title: "Which TV?"
input "tSensor", "capability.temperatureMeasurement"
input "threshold1", "number", title: "Higher than?"
input "window1", "capability.switch"
def installed() {
	subscribe(tv1, "switch", onHandler)
}
def updated() {
	unsubscribe()
	subscribe(tv1, "switch", onHandler)
}
def onHandler(evt) {
	def t = tSensor.currentValue("temperature")
	if ((evt.value == "on") && (t > threshold)) turnOnWindow()
}
def turnOnWindow() {
	if(window1.currentSwitch == "off")
		window1.on()
}
\end{lstlisting}

We present the rule representation in Section~\ref{sec:rule-representation} and
then the rule extraction technique in Section~\ref{sec:rule-extraction}.
While our idea is applicable to multiple smart home platforms, 
we concretely demonstrate the proposed techniques on Samsung SmartThings platform.
For the convenience of discussion, we develop 5 SmartApps, \texttt{ComfortTV}, \texttt{ColdDefender}, \texttt{CatchLiveShow}, \texttt{BurglarFinder}, and \texttt{NightCare}, which implement \textbf{Rule 1-5} depicted in Figures \ref{fig_action_conflicts}, \ref{fig_covert_rules} and \ref{fig_condition_interference_threats}, respectively. We use the code of \texttt{ComfortTV} (see Listing~\ref{listing_rule1}) to discuss the rule extraction. 
Discussion about rule extraction on other platforms is in Section~\ref{section_discussion_limitation}.

\subsection{Rule Representation}
\label{sec:rule-representation}
The goal of our rule extraction module is to 
precisely extract rule-related information from the app code
and represent the information in a uniform form.

\begin{lstlisting}[caption={The rule representation format},label=listing_representation,captionpos=t, abovecaptionskip=2pt, numbers=none, linewidth=\linewidth, aboveskip=3mm]
Trigger:
(:subject).(:attribute)
(:constraint)
Condition:
(:data constraints)
(:predicate constraints)
Action:
(:subject)->(:command)(:paras)(:when)(:period)
(:data constraints)
\end{lstlisting}

Listing~\ref{listing_representation} shows the structured rule representation format we use. 
It contains detailed and precise information about 
the \emph{trigger}, \emph{condition}, and \emph{action} of a rule.
(1) the \emph{trigger} is defined in terms of \textbf{subject} (e.g., a certain device), \textbf{attribute},
and \textbf{constraint} (that should be satisfied for executing the rule). 
(2) the \emph{condition} comprises the \textbf{data constraints} (that
describe how variables are assigned values) and the \textbf{predicate constraints}
(that should be satisfied for proceeding to invoke the action).
(3)  the \emph{action} depicts the \textbf{subject} on which \textbf{command} is issued, where
\textbf{paras.} denotes the parameters related to the command and \textbf{data constraints} denotes all quantitative constraints involving the command parameters; plus,
\textbf{when} denotes the scheduled time and \textbf{period} indicates the repetition interval for issuing the command. By default, both \textbf{when} and \textbf{period} are equal to 0, meaning that the command should be issued with no delay and only once, respectively.

\subsection{Symbolic Execution based Rule Extraction}
\label{sec:rule-extraction}
We perform a static analysis on the source code of SmartApps to extract rules. Although most rules defined by SmartApps are static, we also handle rule dynamics due to configuration updating and the dynamic features of Groovy in Section~\ref{section_implementation} and Section~\ref{section_discussion_limitation}, respectively. 

\vspace{3pt}
\noindent\textbf{Why did prior approaches fail?}
Prior approaches~\cite{jia2017contexiot,wang2018fear} instrument SmartApps
to insert runtime logging logic, so that when sensitive commands are
issued during runtime, the \emph{context} information can be collected.
Such runtime logging approaches do not work for our purpose, as
they only provide the information for rules that \emph{have been executed},
while our goal is to extract \emph{all} the rules \emph{before} they are executed.
SmartAuth~\cite{tian2017smartauth} searches the Abstract Syntax Tree (AST) of the SmartApp source code
to look for information of interest (e.g., the trigger event, the attribute, and the action)
\emph{without tracking the data flows}, so it cannot precisely retrieve
the constraint information due to variable assignments and nested branches.

\vspace{3pt}
\noindent\textbf{Why symbolic execution?}
In order to extract the rules of a SmartApp \emph{completely} and \emph{precisely},
we propose to symbolically execute the app, exploring
all of its execution paths. Each path starts from an
entry point and ends at a sensitive command (i.e., sink):
the command reveals the \emph{action} of a rule,
while the path condition exposes the rule \emph{trigger} and
the \emph{condition}.  
In order to enable symbolic execution on SmartApps, 
\emph{the following questions and technical challenges need to be resolved}.

\vspace{3pt}
\noindent\textbf{Path search strategy.}
A well-known limitation about symbolic execution is
its poor scalability due to path explosion. However,
as SmartApps are small and have limited paths, we are able to analyze them
without encountering the path explosion problem. 
A simple \emph{depth-first path search strategy} works
well in our system.

\vspace{3pt}
\noindent\textbf{Symbolic inputs.}
Data whose values are not dependent on other data are handled as \emph{symbolic inputs} or \emph{sources}. 
In SmartApps, sources include device references, device attribute values, device events, user input, HTTP response, 
constant values and return values of several APIs (see \textbf{API modeling} below). We achieve a completely 
automatic symbolic input identification. We parse all 
\texttt{input} method calls to collect device references (each device reference points to a globally unique 128-bit identifier for a home device connected to SmartThings) 
and user inputs (variables whose values are specified by users during app installation or update) and add a \emph{symbolic input} label to each of them. Besides, we define variables to 
denote device attribute values used in the code and label them as symbolic inputs. 
Similarly, variables which accept HTTP responses and constant value are also labeled as sources. 
A special case is \texttt{State} and \texttt{atomicState}, which are objects for storing a small 
amount of data which can be shared across multiple SmartApp executions. We regard them as symbolic inputs as well. For example, in Listing~\ref{listing_rule1}, the devices references (\texttt{tv1}, \texttt{tSensor}, \texttt{window1}),
the user input (\texttt{threshold}), and the return value of the API call at Line 13 are automatically 
labeled as symbolic inputs.

\vspace{3pt}
\noindent\textbf{Analysis entry points and sinks.}
In our implementation, the analysis entry points include the lifecycle methods, \texttt{installed}, \texttt{updated} and \texttt{uninstalled}.
The analysis \emph{sinks} include capability-protected device commands and security sensitive 
SmartThings APIs (such as \texttt{setLocationMode()}). We consider 126 device control commands 
protected by 104 capabilities \cite{smartthings2018capability} and 21 SmartApp APIs (See Appendix~\ref{appendix_smartthings_api_sink}).

\vspace{3pt}
\noindent\textbf{Generating Control-Flow Graph (CFG).}
We follow the approach in \cite{jia2017contexiot} to generate a control-flow graph built on AST transformation. Our design is to model the trigger-condition-action structure of a rule. A rule with a trigger usually starts from an event subscribed by in a \texttt{subscribe(dev,attr,hndl)} method (typically invoked in one of the analysis entry point methods). The \texttt{subscribe()} call means that when the device \texttt{dev}'s attribute \texttt{attr} changes, the event handler \texttt{hndl} should be invoked. Therefore, each \texttt{subscribe} method represents a trigger. Then we trace into the handler to identify sinks along the execution path. The path branches at conditional statement (e.g., \texttt{if} or \texttt{switch} statement) so we may reach different sinks, which are extracted as actions; the boolean expressions within the condition statements along the execution path from an entry point to a sink are used to construct the condition for that sink. The corresponding trigger, condition, and action are assembled into a rule.

\vspace{3pt}
\noindent\textbf{Constraints for the \emph{trigger} and \emph{condition}.}
The \texttt{subscribe} method may define a trigger in different ways, i.e., using a state 
change (e.g., \texttt{subscribe(tv1, "switch", onHandler)}) or a certain value 
(e.g., \texttt{subscribe(tv1, "switch.on", onHandler)} to trigger the execution of the handler. 
If a conditional statement follows along the execution path to compare 
the subscribed event's value (e.g., Line 14 in Listing~\ref{listing_rule1}), the comparison in 
terms of the event's value is regarded as part of the trigger constraint; otherwise, the trigger 
is only a state change and has no constraint.


We track all data constraints and predicate constraints along the execution path from the entry point to 
sinks and attach them (excluding the trigger constraint) to the rule condition.
We establish data constraints from each value assignment statement. Specifically, we write callback methods in the
compiler to handle the 38 expression types defined in Groovy's documentation \cite{groovy}. On the other hand, we also build predicate constraints from condition statements, i.e., each boolean expression in an \texttt{if} statement or each case expression in a \texttt{switch} statement is translated into a constraint. We also handle the \emph{ternary} expressions by breaking each of them into two branches. 

\vspace{3pt}
\noindent\textbf{API modeling.}
The main challenge is to deal with the closed-source APIs provided by SmartThings. 
We first model the 10 SmartApp APIs\footnote{See Appendix~\ref{appendix_smartthings_api_sink} for the details.} that schedule the execution of specified methods 
according to their arguments and functionalities. For example, \texttt{runIn(delay, method)} 
executes \texttt{method} with the specified time delay. We attach the delay information to the 
scheduled method and continue to trace into the scheduled method to identify sinks. 
The successive sinks are also attached with the delay (note that we use a \emph{when} property to handle delayed commands).

To model APIs that may be involved in the constraint construction, we model the objects, methods and 
object property accesses by manually reviewing the SmartThings developer documentation \cite{smartthings2018documentation}. 
The return values of API methods and the object properties that do not rely on other data are also labeled as \emph{symbolic inputs}.
We model 173 API methods and 94 object property accesses in total and rewrite a static modeling function for 
each method or property access according to its arguments and return value. We further model a portion of external 
Java APIs that are used by SmartApps. Based on these modeling functions, we are able to construct constraints over 
expressions that contain API calls.

\vspace{3pt}
\noindent\textbf{Compiler customization.}
To build the symbolic executor, we implement a compilation customizer instance and add it to the compiler configuration, which is supported by Groovy to allow developers to modify the compilation process at a certain phase. We work at the semantic analysis phase where the compiler creates a class node for each element (variable, method, expression, statement) in the source code and a set of visit methods that follow the generic Visitor pattern \cite{palsberg1998essence} can be implemented for different class node types to specify how the compiler processes these nodes.

As a concrete example, Table~\ref{table_representation_instance} shows the result of rule extraction on the code in Listing~\ref{listing_rule1}.

\begin{table}[t]
	\caption{Rule representation of \textbf{Rule 1}.}
	\renewcommand\arraystretch{1.3}
	\newcommand{\tabincell}[2]{\begin{tabular}{@{}#1@{}}#2\end{tabular}}
	\centering
	\footnotesize
	\begin{tabular}{ p{1.9cm}  p{3.5cm}  p{2cm} }
		\toprule
		\small Trigger & \small Condition & \small Action \\
		\midrule
		\tabincell{l}{\textbf{subject}: tv1 \\ \textbf{attribute}: switch \\ \textbf{constraint}: \\tv1.switch==on} 
		& 
		\tabincell{l}{\textbf{data constraints}: \\t = tSensor.temperature\\ tSensor.temperature=\#DevState \\threshold1 = 30 \\ \textbf{predicate constraints}: \\t $>$ threshold1 \\window1.switch == off} 
		&
		\tabincell{l} {\textbf{subject}: window1 \\ \textbf{command}: on \\ \textbf{paras}: [] \\ \textbf{data constraints}:\\ \text{[]} \\ \textbf{when}: 0 \\ \textbf{period}: 0}\\
		\bottomrule		
	\end{tabular}
	\label{table_representation_instance}
	\vspace{2mm}
\end{table}

\section{\tshort Threat Detection}
\label{section_threat_detector}
Whenever a new app is installed or the configuration of an installed app is updated, our \emph{Threat Detector} detects \tshort threats 
by evaluating the interaction relations between the rules of the installed or updated app and those of apps already installed in the smart home.

\vspace{-2mm}
\subsection{Detecting Action-Interference Threats}

We detect the Actuator Race (AR) and Goal Conflict (GC) threat (see Table~\ref{table_attacks}) between two 
rules $R_{1}$ and $R_{2}$ in two steps: action analysis and then overlapping-constraint detection. 

\subsubsection{Action Analysis}
\label{subsection_action_analysis}
To detect Actuator Races, we first examine if the actions of  $R_{1}$ and $R_{2}$ 
issue contradictory commands to the same actuator, or issue the same command with 
contradictory parameters; either of the two situations indicates the rule pair is an Actuator Race candidate. 
We use device IDs from the \emph{configuration recorder} (see Section~\ref{section_implementation}) 
to determine if the two rules 
control the same device. We maintain a global mapping $M_{AR}$ that maps a
device ID to the list of rules that operate on the device along with the
command and parameter information. The mapping will be used in the subsequent
overlapping-condition detection step. 

To detect Goal Conflicts, we perform a goal analysis to determine if two actions contradict over a common goal. 
A goal in smart home consists of many measurable properties, such as temperature, 
illuminance, humidity, noise, etc. 
We consider how these properties are affected by each command of a 
device type. The effects are denoted as 
$+$ (increasing), $-$ (decreasing) and $\#$ (irrelevant). Accordingly, we construct another 
global mapping $M_{GC}$ for the goal analysis. 
Note that virtual actuators (e.g., mode) that have no direct effect on the goal properties 
are not included in $M_{GC}$. $R_1$ and $R_{2}$ are considered as a Goal Conflict 
candidate if their actions have opposite effects on the same goal property. 

\subsubsection{Overlapping-Condition Detection}
To determine whether a candidate rule pair $R_{1}$ and $R_{2}$ can lead to a real threat, we also need to 
know if they can be executed simultaneously in a certain situation, i.e., if they 
have overlapping trigger and/or condition. Note that we establish constraints for trigger and 
condition in our rule extraction (Section~\ref{sec:rule-extraction}). Therefore, the overlapping detection is transformed 
into a \emph{constraint satisfaction} problem. We merge all constraints of the two rules 
as well as additional device constraints\footnote{Device constraints are to determine if two rules use the same device. See Section~\ref{section_implementation} for more information.}; 
if the problem is solvable\footnote{We choose the Java Constraint Programming (JaCoP) 
library as the solver since it is efficient and open-source in our implementation.}, 
it means two rules take effect together under certain situations (which can be derived from
the resolution results), and the rule pair causes a true Action-Interference Threat.

\subsection{Detecting Trigger-Interference Threats}
The detection of Covert Triggering (CT) threat over two rules $R_{1}$ and $R_{2}$ is 
a directed process since Covert Triggering is not mutative. There are two ways 
that $R_{1}$ triggers $R_{2}$: (1) $R_{1}$ issues a command to an actuator (e.g., a switch), 
changing its certain state that is the trigger of $R_{2}$; (2) $R_{1}$ changes an 
environment feature (e.g., temperature) sensed by a sensor device whose reading is used as $R_{2}$'s trigger. 
The trigger checking step follows the above two ways to determine whether $R_{1}$ could trigger $R_{2}$.


If the rule pair passes the trigger checking step, it is a CT candidate and 
an overlapping-condition detection is performed on the 
conditions of $R_{1}$ and $R_{2}$ to see if they are likely to execute together in certain 
situations. If the rule pair is a CT candidate and has overlapping conditions, 
$R_{1}$ and $R_2$ constitute a CT pair, denoted as $CT_{R_{1}\rightarrow R_{2}}$.

When $CT_{R_{1}\rightarrow R_{2}}$ (or $CT_{R_{2}\rightarrow R_{1}}$) is detected, 
we further detect Self Disabling (SD) threats by examining the action analysis 
result in Section~\ref{subsection_action_analysis}. If $R_{1}$ and $R_{2}$ are also an
Actuator Race candidate, an SD threat is detected. In addition, if both $CT_{R_{1}\rightarrow R_{2}}$ 
and $CT_{R_{2}\rightarrow R_{1}}$ are detected and $R_{1}$ and $R_{2}$ are AR 
candidate, a Loop Triggering Interference (LT) threat is discovered.

\subsection{Detecting Condition-Interference Threats}
To detect whether the rule $R_{1}$ has Enabling Condition (EC) Interference and Disabling Condition (DC) Interference with rule $R_{2}$, we evaluate whether
 $R_1$ enables/disables $R_{2}$'s condition. Similar to the detection of Trigger Interference threats, there are two ways that $R_1$
can affect $R_{2}$'s condition: (1) $R_1$ issues a command to an actuator, which changes the satisfaction of $R_2$'s condition
(e.g., $R_1$ turns on a heater and $R_{2}$ checks if the heater's state is \texttt{on});
and (2) $R_1$ changes an environment feature to enable/disable the condition of $R_2$ (e.g., $R_1$ turns on the heater and the condition
of $R_2$ involves the room temperature). If $R_{1}$ affects $R_{2}$ in either of the above ways, the rule pair is a condition-interference threat candidate. 

We then determine whether $R_{1}$ enables or disables $R_{2}$'s condition by another overlapping-condition detection. Specifically, we first create an \emph{effect constraint} to denote the effect of $R_{1}$'s action. For instance, if $R_{1}$'s action locks a door (\texttt{door1}), we generate the constraint \texttt{door1.lock==locked}; if $R_{1}$ sets the heating temperature of a thermostat to a value $\texttt{T}$ and $R_{2}$ uses a temperature sensor (\texttt{tSensor}) in its condition, the effect constraint is \texttt{tSensor.temperature>=T}. We then merge the effect constraint with $R_{2}$'s condition and solve the new constraint satisfaction problem. If the problem is solvable, $R_{2}$'s condition may be enabled and otherwise disabled by $R_{1}$.
%

\subsection{Detecting Chained \tshort Threats} \label{sec:chained}
There may exist apps which satisfy one of the interference patterns but are still installed, decided by users. 
Hence, when we detect a rule $r_{1}$ defined by a new app interferes with (or is interfered with by) an existing rule $r_{2}$, 
we also need to detect if $r_{1}$ interferes with other rules indirectly via $r_{2}$ (or other rules interfere with $r_{1}$ indirectly). 
To this end, we 
record all rule pairs that satisfy certain pattern but are still installed by users in a list $Allowed$ 
in a bottom-up manner (from the user installed the first app for his home). After the pairwise detection between new rules and old rules, we search this detection result and the $Allowed$ list to find long-chained rules.

\section{Configuration Information Collection}
\label{section_implementation}

\noindent\textbf{Challenge.} In Section~\ref{section_rule_extraction}, we track the data flow by constructing data constraints starting from \emph{sources}. Recall that we take as sources the \texttt{input} methods, which are rendered as user-friendly graphical interfaces for users to specify values for variables (referred to as configuration information in this paper) during app installation or updating. To precisely detect \tshort threats, we need to know the values of such sources.
Take the Actuator Race in the category of 
action-interference threats as an example: the two rules $R_1$
and $R_2$ should operate on the same actuator as a precondition to
cause an Actuator Race. Such device binding as well as other configuration 
information (e.g., a user-defined threshold for comparison) cannot be obtained through static
analysis.
Therefore, how to collect the configuration information without modifying
the SmartThings platform becomes a challenge since there are no APIs available for obtaining 
the configuration information from the SmartThings cloud or the companion app. 


\vspace{2pt}
\noindent\textbf{Solution.} 
An ideal solution is to integrate our \emph{threat detector} and \emph{frontend} modules into SmartThings companion app, so that configuration information can be easily collected within the app. However, the companion app is not open-source and we do not assume any modification on the platform's cloud or mobile app for \tool to work. Hence, we propose a practical solution to obtain configuration information, including a code instrumentation technique to collect configuration information from inside an app, and a messaging mechanism for transmitting the collected information to the \tool app. The \tool app implements the \emph{threat detector} and \emph{frontend} modules as another mobile app.  

\subsection{Code Instrumentation}
Code instrumentation has been used in previous research on the SmartThings 
platform~\cite{jia2017contexiot, tian2017smartauth, wang2018fear}. In our case,
the instrumentation is only to gather the configuration information during 
app installation or configuration update, so it introduces a negligible overhead. We implement the SmartApp instrumentation
with a Groovy script. Listing~\ref{lst-patched}, as an example, shows how the app in Listing~\ref{listing_rule1} is instrumented (most of the unaltered lines are omitted). 

The lifecycle method \texttt{updated} is invoked
during app installation or configuration update, so we insert configuration collection logic in \texttt{updated}. The inserted lines are the same for all apps except for 
Lines 8-10, which collect the configuration information specific to each app. 
The \texttt{appname} can be obtained from the metadata in the \texttt{definition} method\footnote{\texttt{Definition} is a method that determines how the SmartApp is described in the mobile app UI.}. 
In each item of the lists \texttt{devices} and \texttt{values}, the \texttt{devRefStr} and \texttt{varStr} are the variable names defined in \texttt{input} methods, and \texttt{devRef} and \texttt{var} denote the real values specified by users in the mobile app. 
The Groovy script reuses some code of the rule extractor to identify the \texttt{appname}, \texttt{devRefStr}, and \texttt{varStr}, so that the code instrumentation is a completely automatic process. 
The \texttt{collectConfigInfo} method (Line 14) assembles a Uniform Resource Identifier \texttt{uri} that contains the app name, a list of mappings between the device variable name \texttt{devRefStr} and the underlying unique 128-bit device ID (\texttt{devRef.getId()} in Line 17), and the variable names and values (Line 20). 


\begin{lstlisting}[caption={Code snippet showing how the app in Listing \ref{listing_rule1} is instrumented. Line 3, Lines 8--11 and Lines 14--23 are the inserted code.},label=lst-patched, float=tp,floatplacement=tbp, abovecaptionskip=0pt, basicstyle=\scriptsize]
//...
//Specify the phone that runs HomeGuard
input "patchedphone", "phone", required: true, title: "Phone number?"
//...
def updated() {
	//...
	// inserted code
	def appname = "ComfortTV"
	def devices = [[devRefStr:"tv1", devRef:tv1], [devRefStr:"tSensor", devRef:tSensor], [devRefStr:"window1", devRef:window1]]
	def values = [[varStr:"threshold1", var:threshold1]]
	collectConfigInfo(appname, devices, values)
}
//...
def collectConfigInfo(appname, devices, values) {
	def uri = "http://my.com/appname:${appname}/"
	devices.each { dev ->
		uri = uri + dev.devRefStr + ":" + dev.devRef.getId() + "/"
	}
	values.each { val->
		uri = uri + val.varStr + ":" +val.var + "/"
	}
	sendSmsMessage(patchedphone,uri)
}
\end{lstlisting}

\subsection{Messaging Deployment}
We have two options for sending the above URI from SmartThings cloud\footnote{Note that SmartApps run on SmartThings backend cloud.} to the \tool app: SMS (short messaging service) and HTTP based messaging. Both approaches are widely used for messaging and have their own pros and cons. For instance, SMS is easy to deploy but may fail to work if users go abroad, while HTTP works internationally but requires a relay server due to mobile IP addressing issues. We implement both SMS and HTTP solutions in our proof-of-concept prototype.

\begin{figure}
	\centering
	\subfigure[The unique 128-bit device IDs and static values are encapsulated in a URI.]{
		\label{fig_view} 
		\includegraphics[width=0.19\textwidth]{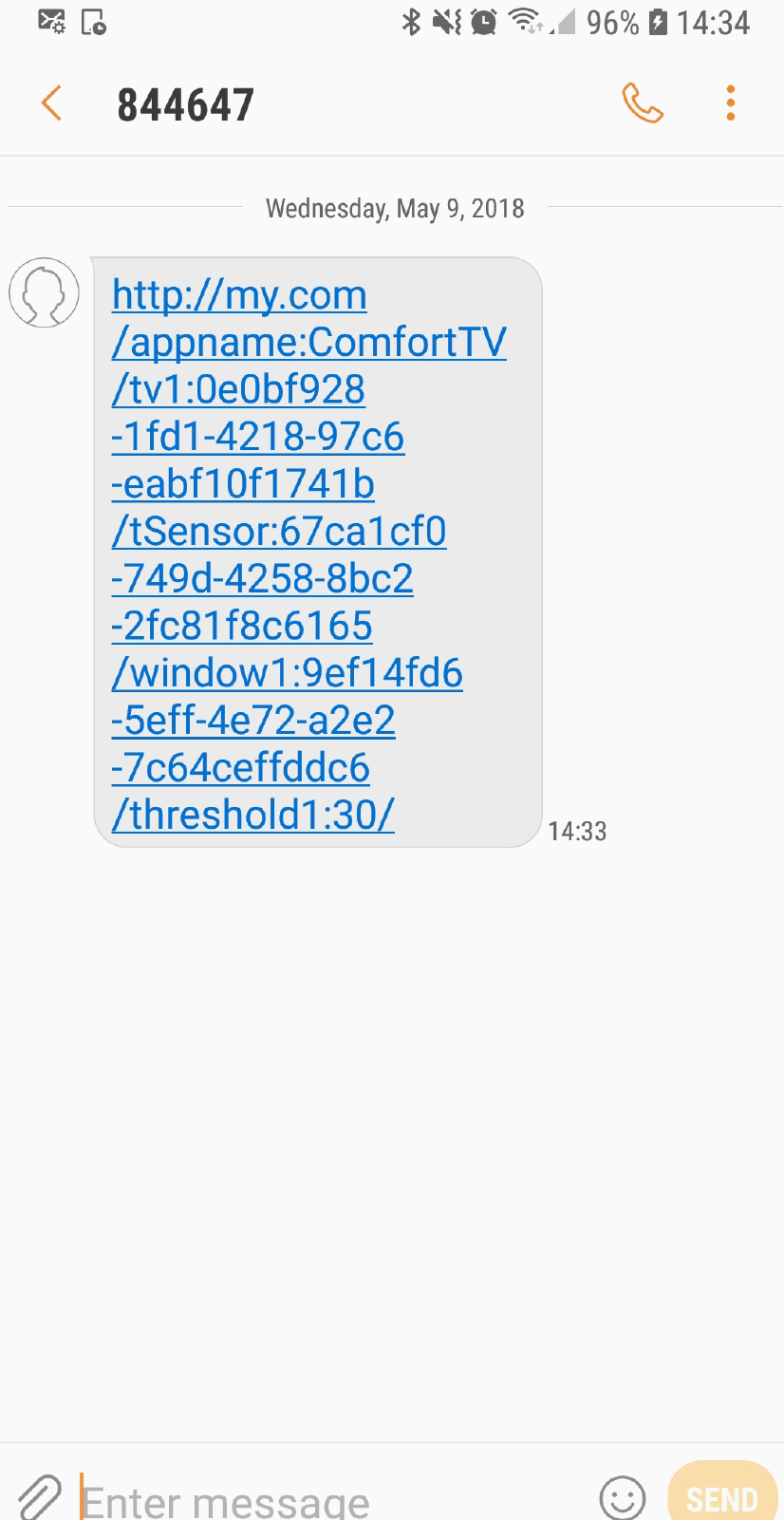}}
	\hspace{1em}
	\subfigure[The frontend app shows the rule defined by \texttt{ComfortTV} and the detected \tshort threats.]{
		\label{fig_prd} 
		\includegraphics[width=0.19\textwidth]{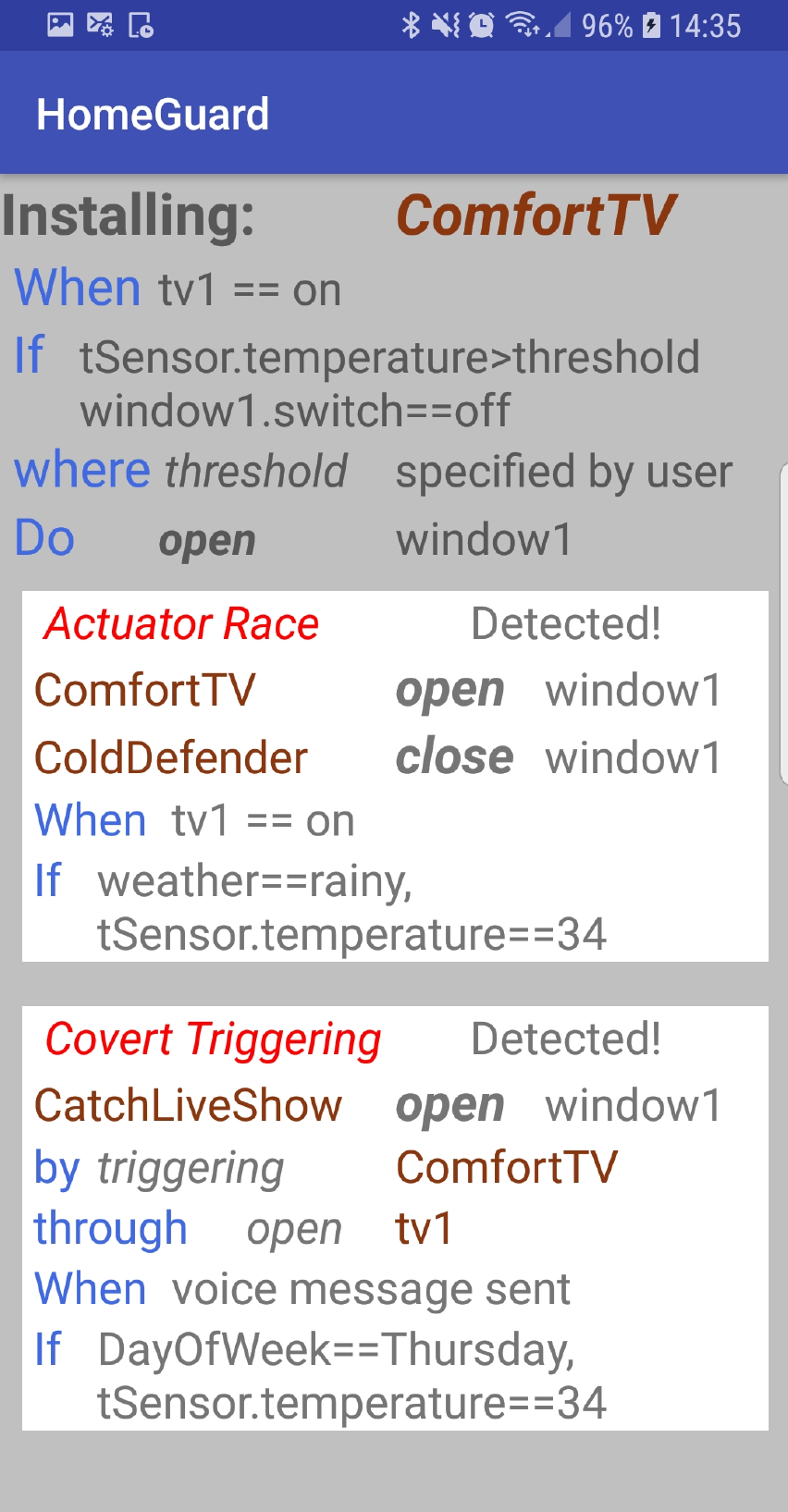}}
	\caption{Screenshots showing the collected configuration information and the \tool frontend 
		app interface when installing \texttt{ComfortTV} to a home with \texttt{ColdDefender} and \texttt{CatchLiveShow} already installed. }
	\label{fig_secrecy} 
	\vspace{-1.5em}
\end{figure}

SMS messaging is easier to implement since SmartThings provide a convenient API \texttt{sendSmsMessage} for sending messages to a specified phone number. In Listing~\ref{listing_rule1}, Line 3 will be rendered as a UI for homeowners to fill out the phone number and Line 22 sends the collected configuration information (i.e., \texttt{uri}) to the specified phone. When installing or updating an app, the homeowner receives an SMS message containing \texttt{uri}. Fig.~\ref{fig_secrecy}(a) shows an example of the SMS message received by the homeowner's smartphone. 
When the received link is clicked, the \tool app installed on the same phone is launched and receives the link (e.g., by declaring an \texttt{Intent Filter} in Android OS \cite{android2018intent}); it parses the link to obtain the 
configuration information. The \tool app then sends an HTTP request with the \emph{appname} to our backend
server (hosting the rule extractor module) to retrieve the rules of the app. The backend server maintains a database to store the rules extracted from the public apps in SmartThings app store, in order to speed up the response time (note that the backend server also provides an interface to
extract rules from a custom app on demand). Besides, the \tool app also generates device constraints (e.g., \texttt{tv1=0e0b...741b}) and user-defined value constraints (e.g., \texttt{threshold1=30}) based on the collected configuration information for detecting \tshort threats.
As the \tool app records all the history information of the already installed
apps, the threat detection can be performed efficiently. After that,
the detection result is transformed and presented to the user; Fig.~\ref{fig_secrecy}(b)
shows an example. 
The user then can determine whether to keep or delete the new SmartApp, or change its configuration.

The HTTP-based implementation shares a similar design but employs Firebase Cloud Messaging (FCM) \cite{firebase2018} for relaying the messages from SmartThings cloud to the homeowner's smartphone. At the initial startup of the \tool app, Firebase generates a unique registration token for each \tool app instance. In the instrumentation code, we ask the homeowner to specify the registration token instead of the phone number (Line 3 in Listing~\ref{lst-patched}). Thus, the \texttt{uri} is sent to Firebase via the HTTP API \texttt{httpPost} (instead of \texttt{sendSmsMessage}) and the firebase server then pushes \texttt{uri} as a notification message to the \tool app instance specified by the registration token. The notification message is rendered as a Notification \cite{android2018notification} for the \tool app. By clicking the notification, the \tool app is launched to detect \tshort threats.

\section{Evaluation}
We demonstrate how \tshort threats can be
constructed using some seemingly benign malicious apps and how the threats
can be exploited~\ref{sec:attack-demo}. We then evaluate the effectiveness
and efficiency of  \tool in
Section~\ref{sec:effectiveness} and Section~\ref{sec:efficiency}, respectively.

\subsection{Exploitation Experiments} \label{sec:attack-demo}
As presented in Section~\ref{section_threat_model}, new attack vectors
exploiting \tshort threats arise. In order to demonstrate the exploitation
feasibility, we act as an attacker and use 5 SmartApps we have developed, i.e.,  
\texttt{ComfortTV}, \texttt{ColdDefender}, \texttt{CatchLiveShow}, \texttt{BurglarFinder}, and \texttt{NightCare},
to simulate the exploitation of \tshort threats. Each of these apps seems 
to provide useful functionalities. But when these apps are installed in 
the same home, their interplay introduces \tshort threats and can be exploited.
For instance, \texttt{BurglarFinder} and \texttt{NightCare} cause a condition-interference
threat, and the user may be unaware that her \texttt{BurglarFinder} is
already disabled by \texttt{NightCare}, and the attacker may break in
without triggering alerts. We install all 
the five apps on SmartThings Web IDE and observe that these apps interfere with each other 
through the trigger, condition, or action, as discussed in Section \ref{section_attacks}. 
In this experiment, we prove that SmartThings currently has no mechanisms to detect or handle \tshort threats, which validate the criticality and urgency of our research.

\begin{table*}[t]
	\centering
	\scriptsize
	\caption{Extracting rules from malicious apps.}
	\renewcommand\arraystretch{1.3}
	\newcommand{\tabincell}[2]{\begin{tabular}{@{}#1@{}}#2\end{tabular}}
	\newcolumntype{P}[1]{>{\centering\arraybackslash}p{#1}}
	\newcolumntype{M}[1]{>{\centering\arraybackslash}m{#1}}
	\begin{tabular}{p{2cm} p{6cm} p{7cm} c}		
		\toprule	
		\textbf{Attack}	& \textbf{Description}  & \textbf{Malicious SmartApp Name} & \textbf{Can handle?} \\
		\midrule 
		Malicious Control & \tabincell{l}{Embed malicious logics beyond app description} & CreatingSeizuresUsingStrobedLight &\CheckmarkBold\\
		Abusing Permission &  Exploit overprivilege to perform attacks & shiqiBatteryMonitor & \CheckmarkBold \\
		Adware & Embed ads into notification messages & HelloHome/CODetector & \CheckmarkBold \\
		Spyware & Leak private information via HTTP/side channel&\tabincell{l}{LockManager/shiqiLightController/DoorLockPinCodeSnooping}   & \CheckmarkBold \\
		Ransomware & Refuse to take actions until user pay money & WaterValve & \CheckmarkBold \\
		Remote Control &  Execute dynamic commands according to HTTP response & SmokeDetector/FireAlarm & \CheckmarkBold \\
		IPC & Malicious apps exchange information by IPC & MaliciousCameraIPC \& PresenceSensor &  \CheckmarkBold \\
		Shadow Payload & Send sensitive information to attacker's encrypted url & AutoCamera2 & \CheckmarkBold \\ 
		Endpoint Attack &  Trigger malicious functions via HTTP requests & \tabincell{l}{BackdoorPinCodeInjection/DisablingVacationMode} & \XSolidBrush \\
		App Update & Edit the original codes after released & BonVoyageRepackaging/PowersOutAlert & \XSolidBrush \\
		\bottomrule
	\end{tabular}
	\label{table_single_attacks}
\end{table*}

\subsection{Effectiveness}
\label{sec:effectiveness}
\vspace{3pt}
\noindent\textbf{Rule Extraction.}
We first evaluate the implemented \emph{rule extractor}'s ability to 
extract automation rules from SmartApps. We collect all the 182 SmartApps
in the public repository \cite{smartthings2018public} and manually remove the 
36 Web Services SmartApps, as these web services SmartApps expose web endpoints for 
external applications to get device information or control devices through web API calls 
and do not define automation rules themselves \cite{smartthings2018documentation}. 
We manually review the code of the remaining 146 SmartApps and record the rules 
per app. To avoid human errors, we also install these apps and use the simulated devices 
provided by SmartThings to verify the correctness. The manual analysis results are used 
as \emph{ground truth}. Then we use our rule extractor to automatically extract rules 
from these apps and compare the results with the ground truth. 

Our rule extractor can analyze most apps (124 out of 146) correctly. There were 
several special cases we did not expect. \texttt{Feed My Pet} uses 
\texttt{device.petfeedershield} in the \texttt{input} method instead of a capability; 
\texttt{Sleepy Time} uses \texttt{device.jawboneUser}; and \texttt{Camera Power Scheduler}
uses a public API \texttt{runDaily}, which is not documented by SmartThings. We have 
added the non-standard device types into the capability list and modeled 
the undocumented APIs we encountered to fix the issue. 

As our rule extractor employs symbolic execution to explore all the paths in apps systematically, 
it is able to extract the rules in an app precisely and completely.
Thus, we envision that it can help the SmartThings staff review the code of SmartApps to check whether 
a SmartApp contains malicious behaviors. To evaluate its effectiveness in extracting 
rules from malware, which may hide malicious logics and thus impose extra challenges to our rule extractor, 
we ran it over 18 representative malicious 
SmartApps collected from Literatures \cite{fernandes2016security, jia2017contexiot, tian2017smartauth,
wang2018fear} to test if it can extract rules correctly. As shown in Table \ref{table_single_attacks}, 
the rule extractor can obtain the precise rules for the majority of the cases. 
Therefore, the rule extractor can be applied to better detect malicious apps.

Two exceptions are the \emph{endpoint attack} and the \emph{app update attack}. The SmartApps 
used for endpoint attacks are web service apps which do not define automation themselves but expose 
callable endpoints for third parties, i.e., the automation rules are defined outside of SmartApps, 
so the rule extractor cannot obtain the complete automation logics by only analyzing SmartApps. 
However, the rule extractor can still identify malicious logics embedded in the request handler 
methods. The app update attack is difficult to detect by static methods since SmartApp developers 
can update the cloud instances for all users without any user awareness. This can be solved through 
the platform by enforcing the checking whenever an app is updated.

\vspace{3pt}
\noindent\textbf{\tshort Threat Detection on Real Cases.}
In order to demonstrate the capability of \tool in finding \tshort threats in real-world cases, we pick up 90 out of the 146 apps in the SmartThings app repository for testing.
We exclude 56 apps because their functionalities are to send notifications to the home owner's smartphone and do not control devices. As this test is to find all possible pairs having \tshort threats from a pool of apps and it is infeasible to exhaustively try all device-app binding situations, we consider two rules use the same device if they use devices of the same type\footnote{The device types used by a SmartApp can be determined by examining their associated  \texttt{capabilities}.}. To avoid excessive false positives due to this setting, we classify devices 
using \texttt{capability.switch} into different types according to the app description, 
since various types of devices support \texttt{capability.switch} to indicate \texttt{on/off} states. Note that in the real deployment, \tool distinguish if two rules operate on the same device according to the 128-bit device IDs in the collected configuration information (see Section~\ref{section_implementation}). 
We perform the \tshort threat detection for each pair of the 90 apps and record the results. The SmartThings platform provides a well-built simulator and simulated devices, allowing us to verify the correctness of the discovered threats. We also purchase some real devices, including a SmartThings hub v2, a motion sensor, a presence sensor, a multipurpose sensor (combining a contact sensor and a temperature sensor), two bulbs, two smart outlets to reproduce and verify a subset of the discovered threats.

\begin{figure}[t]
	\centering
	\includegraphics[width=0.32\textwidth]{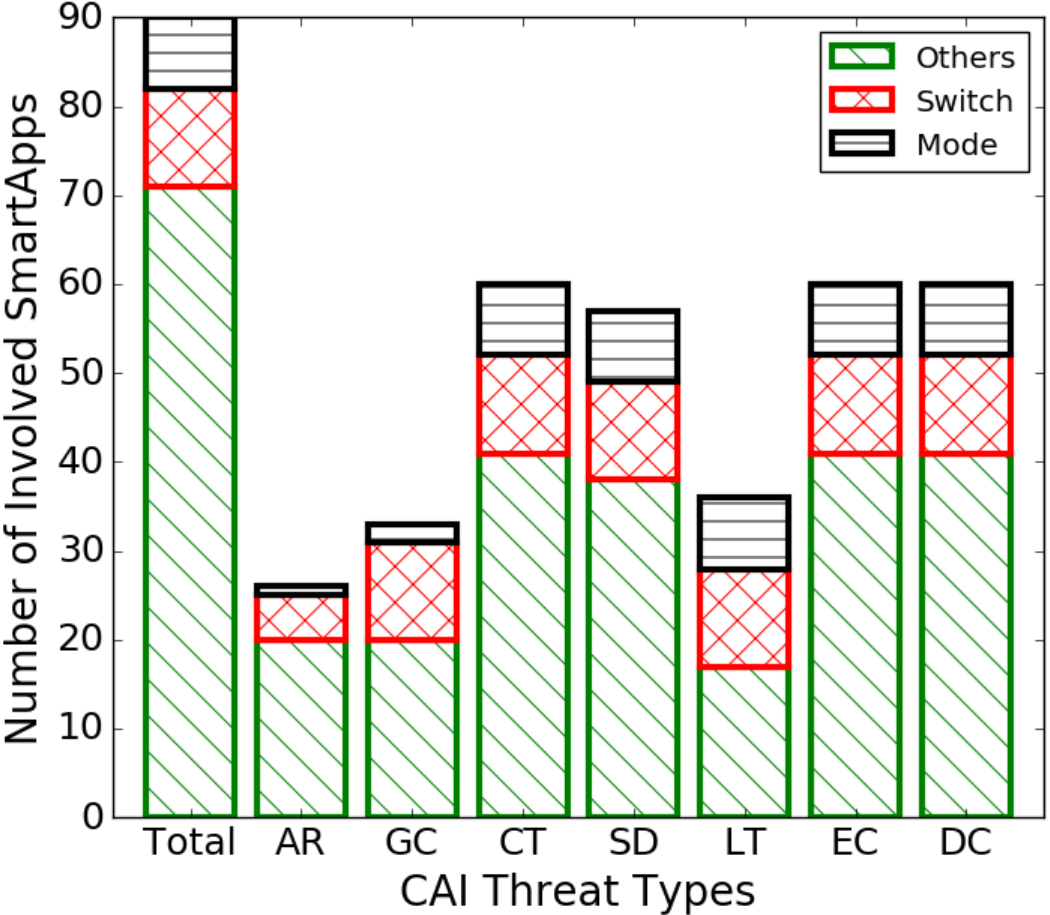}\\
	\caption{Statistics of the detection result on 90 SmartApps. 
		\emph{Switch}: controlling a \texttt{capability.switch} without specifying the specific device type (e.g., light, valve, television, etc.); 
		\emph{Mode}: controlling location \texttt{mode}; 
		\emph{Others}: controlling other devices. 
		The threat acronyms are defined in Table~\ref{table_attacks}.}\label{fig_involved}
	\vspace{-1em}
\end{figure}

We find that a lot of apps can cause \tshort threats and show the statistics 
in Fig.~\ref{fig_involved}. Apps that control a commonly used switch or a mode tend to 
involve all kind of the threats. Below, we describe some detected \tshort threats in SmartApps.
\begin{enumerate}[leftmargin=*]
	\item \texttt{SwitchChangesMode}\footnote{The spaces in SmartApp names are omitted.} changes the current mode of a smart home according to 
	the \texttt{on/off} state of a switch, 
	and \texttt{MakeItSo} binds a group of states of several switches, locks, and 
	thermostats to a mode and restores the group of states each time the home goes into that mode. The two apps 
	may create a covert rule that a switch's state triggers the action of unlocking a door. 
	
	\item \texttt{CurlingIron}, which turns on a set of outlets (switches) if a motion is detected, may create a covert rule 
	by chaining with \texttt{SwitchChangesMode} and \texttt{MakeItSo}. This covert rule unlocks a door 
	when motion is detected. The covert rule introduces a new attack surface that a burglar unlocks the door by spoofing the motion sensor (e.g., using CO2 laser \cite{schleijpen2016using}).
	 
	\item \texttt{NFCTagToggle} allows a user to toggle a set of appliances and door locks by tapping a smartphone app button, and
	\texttt{LockItWhenILeave} locks doors automatically if the user's presence sensor leaves a location. If the user taps the app button 
	to turn off appliances and lock doors after leaving home but \texttt{LockItWhenILeave} has already locked the door, the door will be unlocked. 
	
	\item \texttt{LetThereBeDark} leads to action races on the same lights when working with other light control apps, such as \texttt{UndeadEarlyWarning}, \texttt{LightsOffWhenClosed}, \texttt{SmartNightlight}, \texttt{TurnItOnFor5Minutes}, etc. 
	
	\item \texttt{It'sTooHot} and \texttt{EnergySaver} impose a Self-Disabling threat. \texttt{EnergySaver} turns off a set of devices when the real-time electricity usage is over the user-defined threshold, which may disable \texttt{It'sTooHot} to turn on air conditioners or fans in a narrow situation: the turning-on of air conditioners is the last straw that makes the electricity usage exceed the threshold.
	
	\item \texttt{LightUptheNight} contains a Loop-Triggering interference, leading to unexpected light flashing. This app is a real-world case of the LT example in Section~\ref{section_trigger_interference}.

\end{enumerate}

\subsection{Efficiency}
\label{sec:efficiency}

\vspace{3pt}
\noindent\textbf{Rule Extraction Computation and Storage.}
To test the efficiency of our rule extractor, we run it 10 times on all 146 apps and get the average 
execution time. The time is 1341ms per app on average on a desktop with 3.4GHz Intel 
Core i7 CPU-6700, 8GB memory, and Ubuntu 16.04 LTS. Note that the rule extraction for the publicly 
available apps is a one-time cost and can be done offline. The performance is satisfactory even for 
providing online services to users who would like to install their own custom apps. We also test the 
size of the rule file of a SmartApp, which needs to be stored on the \tool server hosted by \emph{rule 
extractor} and transmitted to the user's mobile device. In our implementation, we use JSON strings 
to store rules and the rule file for an app is 6.2KB on average.

\vspace{3pt}
\noindent\textbf{Configuration Information Collection Speed.}
Both configuration information collection and threat detection are done online and 
their speed directly affects user experience. In our implementation, the latency introduced by configuration information collection depends on the response time of SmartThings cloud and 
the SMS/HTTP transmission latency. We time the response time by inserting invoking (\texttt{now()}) to get the Unix epochs $T_{1}$ and $T_{2}$ before and after the instrumentation code, respectively. The time duration ($T_{2}-T_{1}$) is 27 ms. Similarly, we record the epoch $T_{3}$ when we receive the SMS/HTTP message on the phone and calculate the transmission latency ($T_{3}-T_{2}$). The averaged latency from 100 trials is 3120ms for SMS and 1058ms for HTTP, which we believe is acceptable during installation.

\begin{figure}[t]
	\centering
	\includegraphics[width=0.338\textwidth]{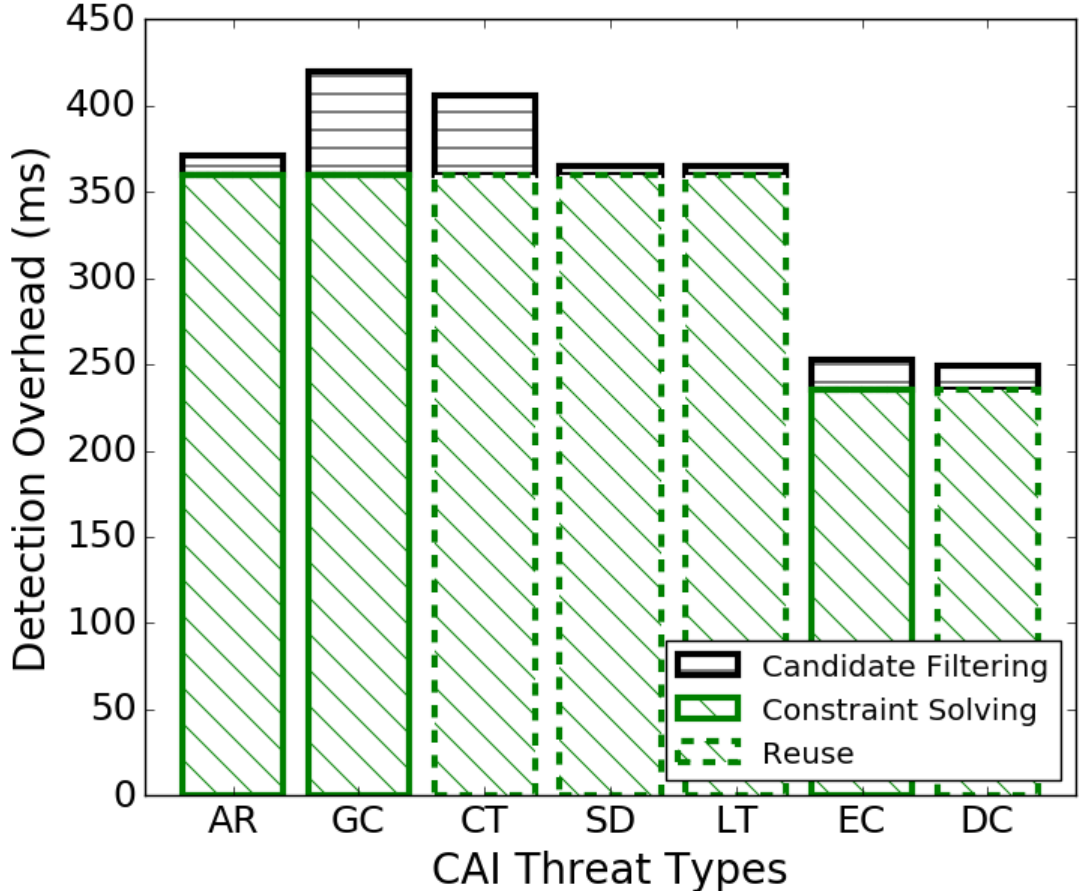}\\
	\caption{\tshort detection overhead for a pair of rules. Green dotted lines mean the constraint solving for detecting CT, SD, and LT threats can reuse the solving result of AR and the constraint solving for DC can reuse that of EC. The threat acronyms are defined in Table~\ref{table_attacks}.}\label{fig_detection_overhead}
	\vspace{-1em}
\end{figure}

\vspace{3pt}
\noindent\textbf{\tshort Detection Speed.}
\tshort detection is a real-time process and runs on users' mobile devices. To evaluate 
the efficiency of \tool, we test the averaged execution time for detecting a specific \tshort 
threat between two rules on a Samsung Galaxy S8 smartphone with Android OS 8.1.0. As shown 
in Fig.~\ref{fig_detection_overhead}, the most time-consuming operation is constraint solving. 
The constraints solving overhead in EC is lower since involved constraint number is half of that in AR and GC. 
To avoid unnecessary constraint solving, we first perform a light-weight candidate filtering based 
on the pre-stored mapping lists and reuse the constraint solving result in detecting different threats. 
For an arbitrary pair of rules, the maximum total time for detecting all \tshort threats is 1156 ms. The actual detection time is much less since two rules rarely fit all threat patterns and thus constraint solving overhead can be avoided.

\subsection{Discussion and Limitations}
\label{section_discussion_limitation}
\subsubsection{User Intervention}
Due to the inherent weakness of dynamic logging, previous work \cite{jia2017contexiot} needs multiple occurrences of user intervention when an app issues a command. By utilizing the powerful path searching capability of symbolic execution, \tool extracts the rules of an app all at once. Hence, users in this paper only participate in a one-time decision making on the \tool frontend app when they manually install a new app or update the configuration of an already installed app. Users focus more on handling app functionalities during installation than during the app's daily execution, so \tool does not risk minimum user habituation or annoyance. It is infeasible to further eliminate the one-time decision making because inter-app interactions are somewhat subjective, as we discussed in Section~\ref{section_goal_scope}.

\subsubsection{Dynamic Features of Programming Languages}
A concern may be that our static symbolic execution engine cannot effectively deal with the dynamic features of Groovy. However, all SmartApps are instances of an abstract class \texttt{Executor} which provides a variety of methods available for app development and run in the sandboxed environment, which is customized and enforced by SmartThings. SmartThings enforces all SmartApps to implement an abstract class \texttt{Executor} which provides a set of methods and restricts access to many Groovy methods and features. Thus, only \texttt{GString} is an allowable dynamic feature in SmartApps. \texttt{GString} enables remote servers to control the property accesses and method calls in a SmartApp by manipulating the string values. However, this concern is relieved because the SmartThings code review bans dynamic method execution and needs developers to use a \texttt{switch} statement on all possible \texttt{GString} values before doing anything with it \cite{smartthings2018codereview}. Therefore, our static analysis handles all possible values of \texttt{GString} separately and branches the execution path when encountering a switch statement. 

\subsubsection{Backward Compatibility}
A practical concern is how to detect \tshort threats in apps that were installed \emph{before} the employment of the proposed system. This problem can be solved by \tool conveniently without developer efforts. Users can reinstall the instrumented versions of these apps on the phone companion app without changing their configurations; the instrumentation code in \texttt{updated} methods will be invoked and then the \tool app starts to detect \tshort threats, as we discussed in Section~\ref{section_implementation}.   

\subsubsection{Multi-Platform Applicability}

\tool supports different IoT platforms by design, considering home members may work with multiple platforms simultaneously. The rule extractor module is platform-specific since platforms may use different languages and customized APIs, as shown in Table~\ref{table_defining_rules}. Our work shows that symbolic execution works well in extracting automation rules from the source code of Groovy-based SmartApps. Symbolic execution has been widely used for software testing in various systems \cite{godefroid2005dart, mirzaei2012testing} and shows its powerfulness for supporting multiple languages such as source code \cite{cadar2008klee}, bytecode \cite{puasuareanu2010symbolic} and custom intermediate representations (IRs) \cite{fratantonio2016triggerscope}. Therefore, the only engineering effort for supporting multiple platforms is to implement our symbolic execution based rule extractor for other languages. Other than programs/apps, some platforms define rules through templates. For example, IFTTT provides graphical interfaces on its mobile app and web page for users to define automation rules by selecting pre-defined templates and filling out parameters. Rules can be extracted by crawling text data on the related pages and parse the texts with natural language processing (NLP) technologies \cite{hwang2016data}.

\begin{table}
	\vspace{-1.5mm}
	\centering
	\footnotesize
	\caption{Manners for defining rules on different platforms}
	\renewcommand\arraystretch{1.2}
	\newcommand{\tabincell}[2]{\begin{tabular}{@{}#1@{}}#2\end{tabular}}
	\newcolumntype{P}[1]{>{\centering\arraybackslash}p{#1}}
	\newcolumntype{M}[1]{>{\centering\arraybackslash}m{#1}}
	\begin{tabular}{M{1.9cm} P{1.2cm} P{2.3cm} M{1.7cm}}		
		\toprule
		\textbf{Platform} & \textbf{Manner} & \textbf{Language} & \textbf{Specific APIs?}\\ 
		\midrule
		Android Things & program & Java & \CheckmarkBold     \\ 
		HomeKit	& program & Swift/Objective C & \CheckmarkBold \\ 
		OpenHAB & program & Domain Specific Language & \CheckmarkBold \\ 
		SmartThings	& program & Groovy & \CheckmarkBold \\ 
		IFTTT & template & -- & -- \\ 
		\bottomrule
	\end{tabular}
	\label{table_defining_rules}
\end{table}

\section{Related Work}
\label{section_related_work}

 \begin{table} 
 	\centering
 	\caption{Comparison with related work.}
 	\footnotesize
 	\newcommand{\tabincell}[2]{\begin{tabular}{@{}#1@{}}#2\end{tabular}}
 	\begin{tabular}{  c  c  c  c  c  }
 		\toprule
 		\textbf{Name}  & \textbf{\tabincell{c}{Inter-app\\ Analysis}} & \textbf{\tabincell{c}{Proactive\\Defense}}  & \textbf{\tabincell{c}{Low\\ Overhead}} & \textbf{\tabincell{c}{No Runtime\\ Intervention}} \\
 		\midrule
 		ContexIoT   &    \XSolidBrush & \XSolidBrush & \XSolidBrush & \XSolidBrush \\
 		ProvThings   &  \CheckmarkBold    & \XSolidBrush & \XSolidBrush &  \CheckmarkBold \\
 		SmartAuth   &   \XSolidBrush &  \CheckmarkBold &  \CheckmarkBold &  \CheckmarkBold \\
 		\textbf{\tool}  &  \CheckmarkBold &   \CheckmarkBold &  \CheckmarkBold &  \CheckmarkBold \\
 		\bottomrule
 	\end{tabular}
 	\label{table_comparison}
 \end{table}

\subsection{IoT Security and Privacy}

Recently, IoT platform security has been extensively studied. Fernandes et al. use a black-boxed static analysis on Samsung SmartThings, revealing several significant design flaws, such as coarse capability, coarse SmartApp-SmartDevice binding, and insufficient event data protection, which can or have been exploited by SmartApps to perform overpriviledged actions \cite{fernandes2016security}. ContexIoT proposes a context-based permission system for IoT platforms to identify fine-grained context information and prompt the information at runtime for users to make an access control decision \cite{jia2017contexiot}. To separation the execution of a SmartApp into context collection and permission granting phases, they are the first to use patching mechanism to collect runtime data and pause the execution of SmartApps. However, ContexIoT needs users to participate in making decisions at runtime, which overshadows the benefit of home automation, and worse, may violate the 20-second execution time limits for each method in SmartThings if users do not respond in time. SmartAuth performs static analysis of the source code and uses NLP techniques to analyze code annotations, capability requests of SmartApps and developers' descriptions in the app store, to detect whether a SmartApp's functionalities faithfully follow the expectation of users \cite{tian2017smartauth}. Due to the nature of NLP, SmartAuth suffers if a malicious app uses customized meaningless or even malicious method and property names to hide the real SmartThings commands and attributes. Also, the static analysis for extracting rules of a SmartApp can be improved since the authors only target a subset of data types. Wang et al. presents \emph{ProvThings} \cite{wang2018fear}, a platform-centric logging framework that can construct data provenance graphs for all activities in an IoT system and use them to find out reasons for troubleshooting when an abnormality occurs. The ProvThings is prototyped for the SmartThings by instrumenting SmartApps. ProvThings is mainly for forensics rather than proactive defense. Tyche introduces a risk-based permission system to solve the overprivilege problems by grouping permissions according to their risk levels and allows users to grant permissions for a device by specifying an allowed risk level \cite{rahmati2018tyche}. Tyche enforces the risk-based model by rewriting SmartApps based on AST transformations. FlowFence protects IoT data from leakage and misuse by using sandboxes and taint-tracking to enforce data flows between data sources and data sinks. The implementation is done on an Android OS which is not the mainstream in current IoT hubs and clouds. A comparison of existing appified IoT security solutions is shown in Table \ref{table_comparison}.

In addition to platforms, IoT security and privacy in other dimensions also have been widely studied. Fernandes et al. shed light on the distinctive characteristics of IoT security from classic IT security in hardware, software, network, and application layers \cite{fernandes2017internet}. Arias et al. \cite{arias2015privacy} and Liu et al. \cite{liu2017smart} conduct broad studies on the security threats in IoT device development and deployment. Tan et al. \cite{tan2017mtra} propose an attestation solution to ensure the device firmware and software integrity. Chen et al. propose a fuzzing-based detection framework to find memory corruption vulnerabilities in IoT devices \cite{cheniotfuzzer}. A lot of work focus on identifying cyber attacks \cite{antonakakis2017understanding, siby2017iotscanner} and hardening communication and authentication protocols \cite{zhang2017proximity, gong2017piano, anand2016vibreaker}. Siby et al. design a framework \emph{IoTScanner} for passively monitoring and analyzing IoT traffics \cite{siby2017iotscanner}. Also, formal analytics based on probabilistic models are employed to estimate the security risks and guide the configurations of IoT systems \cite{mohsin2016iotsat, mohsin2017iotchecker}.

\subsection{Collusive Attacks in Mobile Apps}
We discuss collusion attacks in the mobile application domain since collusion attacks share similarities with \tshort threats. That is, collusion attacks also leverage multiple apps to bypass malicious code anti-malware techniques that detect per single app, such as Google Play Protect which is based on machine learning and anti-malware apps (e.g., Avast).

A lot of work have been done to characterize and detect mobile app collusion attacks. Davi et al. \cite{davi2010privilege} shed light upon possible privilege escalation attacks in Android. Xu et al. \cite{xu2017appholmes} study app collusion where one app surreptitiously launches others in background and develop a static analysis tool for detecting app collusion by examining app binaries. Marforio et al. \cite{marforio2012analysis} give a comprehensive introduction of possible collusion channels. Chin et al. \cite{chin2011analyzing} firstly present a comprehensive analysis of threats based on inter-app ICC (Inter-Component Communication). Many collusion app detection works are based on examining inter-app ICC data flows \cite{liu2017mr, klieber2014android, ravitch2014multi, li2015iccta}. Bosu et al. \cite{bosu2017collusive} enhances the scalability and accuracy for large-scale detection by designing a new open-source resolution tool for inspecting inter-app ICC data flows.

However, the collusion attacks and \tshort threats are different in many aspects. Collusion attacks need two or more malicious apps to actively collude by design to launch attacks. That is, a privileged app collects sensitive information and sends it to another app (or apps), which then sends the information outside the boundaries of the device to steal private data, or perform harmful actions (e.g., financial transactions) with the received information. Therefore, explicit collusion logics and inter-app communication supported by the mobile system architecture are necessities for apps to collude. Thus, code analysis and information tracking (e.g., taint analysis \cite{arzt2014flowdroid, enck2014taintdroid}) techniques are effective to identify collusions by focusing on inter-app communication channels. In \tshort threats, apps do not have explicit malicious code for collusion and do not need to convey data. The interplay between apps is through controlling devices and affecting the home environment, which needs to be handled by totally different techniques. Moreover, collusion attacks usually abuse permission to, for example, collect and disclose private data, while \tshort threats cause various conflicts and chained execution without violating or abusing permissions. Last but not least, collusive attacks are achieved via deliberately distributed malicious mobile apps or SDKs (Software Development Kits); other than malicious apps, \tshort threats can also be caused due to installing and/or configuring benign apps.

\section{Conclusion}
\label{sec:conclusion}

In an appified smart home, multiple independently developed apps may
interplay and interfere with each other, causing undesired and even dangerous conflicts
and covert rules, which we call \tlong (\tshort) threats. 
Such threats may not only lead to unexpected automation, but also introduce
security and privacy problems to the smart home. 
We have categorized \tshort threats and introduced new attack vectors
that exploit them. Without proper handling, the problem will exacerbate when
an increasing number of smart devices and apps are installed at a smart home.
We have designed and built a system \tool to address the problem.
It applies symbolic execution to extract rules from apps completely and
precisely and employs a constraint solver to evaluate the relation
between rules for systematic threat detection. Moreover, we have proposed a practical
deployment path that utilizes code instrumentation to collect the
installation information and a frontend app to perform the detection
on the user's smartphone. We evaluated \tool using real SmartApps in the app store and
discovered a large number of potential threats. The evaluation
results show that \tool is effective, efficient and precise.



\bibliographystyle{IEEEtranS}
\bibliography{reference}

%

  \appendix

\subsection{Sinks in the Symbolic Executor}

\label{appendix_smartthings_api_sink}
\emph{Sinks} are defined for the symbolic executor to identify rule actions. The most common \emph{sinks} are commands supported by \emph{capabilities}. Capabilities, similar to \emph{permissions} in mobile applications, abstract various types or subtypes of real devices according to functionalities. A real device supports one or more capabilities and can be granted to SmartApps which request the supported capabilities through \texttt{input} methods. Each capability contains a set of \emph{attributes} that can be accessed by apps and \emph{commands} that control the real device. For example, \texttt{capability.lock} defines an attribute \texttt{lock} which indicates a lock device's \texttt{locked}/\texttt{unlocked} state and two commands \texttt{lock()} and \texttt{unlock()} for controlling a lock.  The capability-defined commands are considered as sinks.
 
In addition, SmartThings provide a lot of APIs for building SmartApps and device handlers. To filter out irrelevant method calls that are not home automation actions, we only consider APIs that perform sensitive actions (see Table~\ref{table_apis}), e.g., read attributes from sensors, issue commands to actuators, or send data to third-party devices or servers, as sinks (i.e., rule actions) in our rule extraction.  

\begin{table}[ht!]
	\centering
	\scriptsize
	\caption{SmartThings provided APIs we considered as \emph{sinks} in rule extraction.}
	\renewcommand\arraystretch{1.178}
	\newcommand{\tabincell}[2]{\begin{tabular}{@{}#1@{}}#2\end{tabular}}
	\newcolumntype{P}[1]{>{\centering\arraybackslash}p{#1}}
	\newcolumntype{M}[1]{>{\centering\arraybackslash}m{#1}}
	\begin{tabular}{|p{2.4cm}|p{5.6cm}|}		
		\hline
		\footnotesize \textbf{API Name} & \footnotesize \textbf{Description} \\
		\hline
		\tabincell{l}{
			\texttt{httpDelete} \\
			\texttt{httpGet} \\
			\texttt{httpHead} \\
			\texttt{httpPost} \\
			\texttt{httpPostJson} \\
			\texttt{httpPut} \\
			\texttt{httpPutJson} \\
		} &  Executes an HTTP DELETE/GET/HEAD/POST/PUT request \\\hline
		\texttt{runIn} & Executes the specified method after the specified seconds \\\hline
		\tabincell{l}{
			\texttt{runEvery1Minute} \\
			\texttt{runEvery5Minutes} \\
			\texttt{runEvery10Minutes} \\
			\texttt{runEvery15Minutes} \\
			\texttt{runEvery30Minutes} \\
			\texttt{runEvery1Hour} \\
			\texttt{runEvery3Hours} \\
		} &  Creates a recurring schedule that executes the specified method periodically, as indicated by the method name \\\hline
		\texttt{runOnce} & Executes the specified method once at the specified date and time \\\hline
		\texttt{schedule} & Creates a scheduled job that calls the specified method once per day at the specified time \\\hline
		\texttt{sendHubCommand} & Sends a command to the SmartThings hub, which then issues a command defined in the arguments to LAN-connected devices to access device attributes or control these devices \\\hline
		\tabincell{l}{
			\texttt{sendSms} \\
			\texttt{sendSmsMessage} \\
		} & Sends an SMS message to the specified phone number \\\hline
		\texttt{setLocationMode} & Set the home's \texttt{Mode} to the specified value, which can be subscribed or accessed by SmartApps \\
		\hline
	\end{tabular}
	\label{table_apis}
\end{table}

\end{document}